\documentclass[preprint,12pt,3p]{elsarticle}

\usepackage{graphicx}
\usepackage{amssymb}
\usepackage{amsmath}
\usepackage{units} 
\usepackage{subfigure}

\usepackage{natbib}
\usepackage{longtable}

\def\nue{$\nu_e$}

\def\B{$\iso{8}{B}$ neutrino}
\def\fB{\Phi_{\rm B}}

\newcommand{\hto}{H$_2$O}
\newcommand{\dto}{D$_2$O}
\newcommand{\teff}{\ensuremath{T_{\rm eff}}}
\newcommand{\ns}{$^{16}$N}
\newcommand{\cf}{$^{252}$Cf}
\newcommand{\nsixtn}{$^{16}$N}
\newcommand{\iso}[2]{{}^{#1}{\rm #2}}

\newcommand{\Peed}{P_{ee}^{\rm d}(E_\nu)}
\newcommand{\Peen}{P_{ee}^{\rm n}(E_\nu)}

\newcommand{\Aee}{A_{ee}(E_\nu)}
\newcommand{\Peea}{c_0}
\newcommand{\Peeb}{c_1}
\newcommand{\Peec}{c_2}
\newcommand{\Aeea}{a_0}
\newcommand{\Aeeb}{a_1}
\newcommand{\chis}{$\chi^2$}

\newcommand{\flux}{\times10^6\,{\rm cm^{-2}s^{-1}}}
\newcommand{\thetaonetwo}{\theta_{12}}
\newcommand{\thetaonethree}{\theta_{13}}
\newcommand{\thetatwothree}{\theta_{23}}
\newcommand{\Dmonetwo}{\Delta m^2_{21}}

\newcommand{\Dmonethree}{\Delta m^2_{31}}
\newcommand{\tanthetaonetwo}{\tan^2\thetaonetwo{}}
\newcommand{\sinthetaonethree}{\sin^2\thetaonethree{}}
\newcommand{\Encd}{E_{\rm NCD}}
\newcommand{\cts}{\cos\theta_\odot}

\newcommand{\be}{\beta_{14}}
\newcommand{\errorStatSys}[3]{#1\pm #2{\rm (stat.)}\pm #3{\rm (syst.)}}

\newcommand{\errorStatASys}[4]{#1\pm #2{\rm (stat.)}^{+#3}_{-#4}{\rm (syst.)}}

\newcommand{\numberSNOBflux}{$(\errorStatASys{5.25}{0.16}{0.11}{0.13})\flux$}

\newcommand{\numberPeea}{$\Peea{} = \errorStatSys{0.317}{0.016}{0.009}$}

\journal{Nuclear Physics B}

\begin{document}

\begin{frontmatter}

\title{The Sudbury Neutrino Observatory}

\address[label1]{Ottawa-Carleton Institute for Physics, Department of Physics, Carleton University, Ottawa, Ontario K1S 5B6, Canada}
\address[label2]{Department of Physics and Astronomy, University of Pennsylvania, Philadelphia, PA 19104, USA}
\address[label3]{Department of Physics, Queen's University, Kingston, Ontario K7L 3N6, Canada}
\address[label4]{Institute for Nuclear and Particle Astrophysics, Nuclear Science Division, Lawrence Berkeley National Laboratory, Berkeley, CA 94720, USA}

\author[label1]{A.~Bellerive}
\ead{alainb@physics.carleton.ca}

\author[label2]{J.R.~Klein}
\ead{jrk@hep.upenn.edu}

\author[label3]{A.B.~McDonald\corref{cor1}}
\cortext[cor1]{Corresponding author}
\ead{art@snolab.ca}

\author[label3]{A.J.~Noble}
\ead{potato@snolab.ca}

\author[label4]{A.W.P.~Poon}
\ead{awpoon@lbl.gov}

\author{for the SNO Collaboration}

\begin{abstract}
This review paper provides a summary of the published results of the
Sudbury Neutrino Observatory (SNO) experiment that was carried out by
an international scientific collaboration with data collected during
the period from 1999 to 2006. By using heavy water as a detection
medium, the SNO experiment demonstrated clearly that solar electron
neutrinos from $^8$B decay in the solar core change into other active
neutrino flavors in transit to Earth. The reaction on deuterium that
has equal sensitivity to all active neutrino flavors also provides a
very accurate measure of the initial solar flux for comparison with
solar models. This review summarizes the results from three phases of
solar neutrino detection as well as other physics results obtained
from analyses of the SNO data.
\end{abstract}


\end{frontmatter}


\section{Introduction}

The Sudbury Neutrino Observatory (SNO) was initiated in 1984 primarily
to provide a definitive answer to the Solar Neutrino
Problem~\cite{jnb_na}. Ever since the pioneering calculations of solar
neutrino fluxes by John Bahcall and the pioneering measurements by Ray
Davis in the 1960's, it was known that there was a discrepancy between
the observed fluxes and the calculations.  The persistence of the
problem motivated Herb Chen to contact Canadian scientist Cliff
Hargrove, a former colleague, to explore whether there was a
possibility that enough heavy water could be made available on loan to
perform a sensitive measurement and determine whether the neutrinos
change their type in transit from the core of the Sun. The unique
properties of deuterium could make it possible to observe both the
electron neutrinos produced in the core of the Sun and the sum of all
neutrino types~\cite{chen}.  With the immediate involvement of George
Ewan, who had been exploring underground sites for future experiments,
a collaboration of 16 Canadian and US scientists was formed in 1984,
led by Chen and Ewan as Co-Spokesmen~\cite{sinclairINC}. UK scientists
joined in 1985, led by David Sinclair as UK Spokesman.

An initial design was developed, to be sited 2 km underground in
Inco's Creighton mine near Sudbury, Ontario, Canada and preliminary
approval was obtained from Atomic Energy of Canada Limited (AECL) for
the loan of 1000 tonnes of heavy water. Unfortunately Herb Chen passed
away tragically from leukemia in 1987.  The collaboration continued
with Art McDonald and Gene Beier as US Spokesmen and grew with the
addition of other institutions in the US and Canada for a total of 13
institutions. In 1989, funding was provided jointly by Canadian, US
and UK agencies and McDonald became Director of the project and the
scientific collaboration.

\section{Science of Solar Neutrinos and Detection by SNO}

\begin{figure}[hbtp]
\begin{center}
\includegraphics[height=0.5\textheight, angle=270]{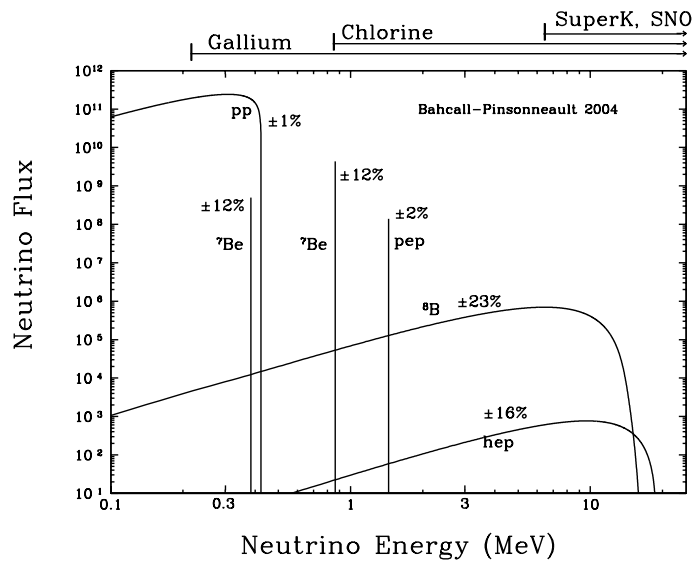}
\caption{Fluxes of neutrinos from the pp chain in the Sun. Threshold energies
for neutrino detection using chlorine, gallium and H$_2$O (Kamiokande
and Super-Kamiokande experiments) are shown. \label{fig:flux}}
\end{center}
\end{figure}

Figure~\ref{fig:flux} shows the fluxes of neutrinos from the pp chain
reactions that comprise the principal power source in the
Sun~\cite{bp2000}.  Overall the series of reactions can be summarized
as:
$4{\rm{p}}  \to {}^4{\rm{He}} + 2{e^ + } + 2{\nu _e} + 26.73{\mbox{ MeV}}$.
Also shown are the thresholds for neutrino detection for the chorine,
gallium and H$_2$O-based experiments that took place before the SNO
results were first reported in 2001. These experiments were either
exclusively (chlorine, gallium) or predominantly (H$_2$O) sensitive to
the electron-type neutrinos produced in the Sun. They all showed
deficits of factors of two to three compared to the fluxes illustrated
in Fig.~\ref{fig:flux}.  It was not possible, however, for these
experiments to show conclusively that this was due to neutrino flavor
change rather than defects in the solar flux calculations.  With heavy
water containing deuterium (\dto), the SNO experiment was able to
measure two separate reactions on deuteron (d):
\begin{enumerate}
    \item ${\nu _e} + {\rm{d}} \to {\rm{p}} + {\rm{p}} + {e^ - }$, a charged current (CC) reaction that was sensitive only to electron-flavor neutrinos, and
    \item ${\nu _x} + {\rm{d}} \to {\rm{n}} + {\rm{p}} + {\nu _x},$ a neutral
current (NC) reaction that was equally sensitive to all neutrino types.
\end{enumerate}

A significant deficit in the $^8$B $\nu$ flux measured by the CC
reaction over that measured by the NC reaction would directly
demonstrate that the Sun's electron neutrinos were changing to one of
the other two types, without reference to solar models. At the same
time, the NC reaction provided a measurement of the total flux of
${}^8{\rm{B}}$ solar neutrinos independent of neutrino flavor change.
The CC reaction was detected by observing the cone of Cherenkov light
produced by the fast moving electron.  The NC reaction was detected in
three different ways in the three phases of the project.  In Phase I,
with pure heavy water in the detector, the NC reaction was observed
via Cherenkov light from conversion of the 6.25-MeV $\gamma$~ray
produced when the free neutron captured on deuterium. In Phase II,
with NaCl dissolved in the heavy water, the neutrons produced via the
NC reaction captured predominantly on chlorine, resulting in a cascade
of $\gamma$~rays with energy totaling 8.6 MeV and producing a very
isotropic distribution of light in the detector. The capture
efficiency was increased significantly during Phase II and the
isotropy enabled a separation of events from the two reactions on a
statistical basis.  In Phase III, the NC neutrons were detected
in an array of ${}^3{\rm{He}}$-filled neutron counters.

In addition, the SNO detector could observe neutrinos of all flavors via the
elastic scattering (ES) of electrons by neutrinos: 
\begin{enumerate}\addtocounter{enumi}{2}
    \item ${\nu _x} + {e^ - } \to {\nu _x} + {e^ - }$ which is six times more sensitive to electron neutrinos than other flavors.
\end{enumerate}
This is the same reaction used by the Kamiokande-II and Super-Kamiokande
experiments to observe solar neutrinos using light water as a medium. 

\section{Experiment Description}

\begin{figure}[htbp]
\begin{center}
\includegraphics[height=0.4\textheight]{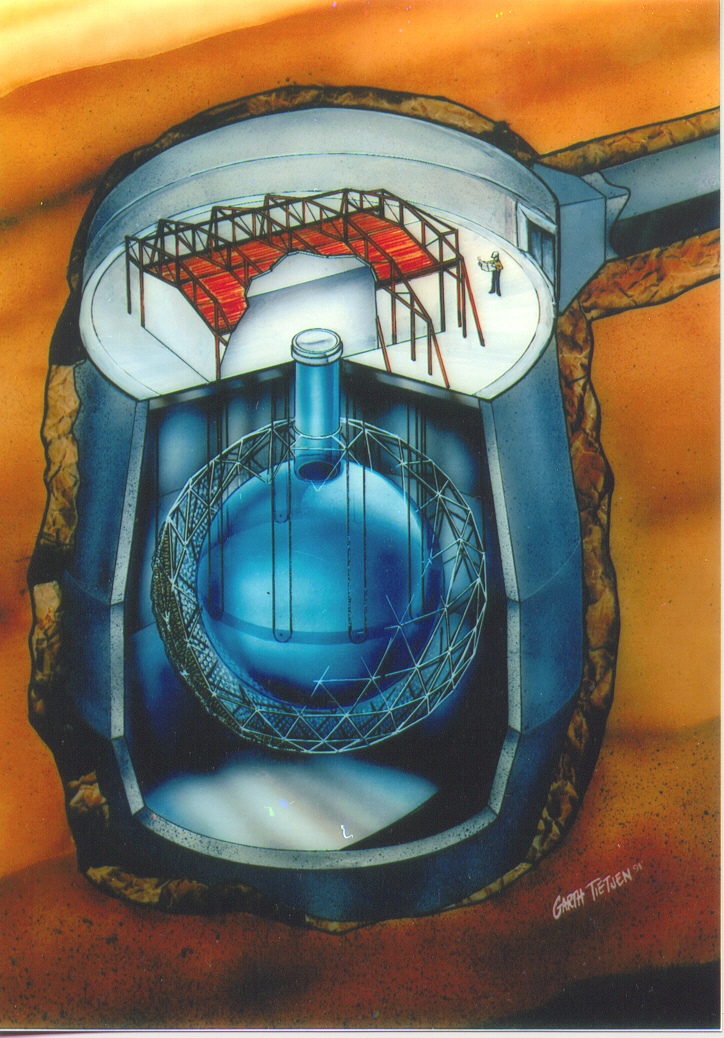}
\caption{Schematic cutaway view of the SNO detector suspended inside the SNO cavity. \label{fig:schematic}}
\end{center}
\end{figure}

Figure~\ref{fig:schematic} is a schematic diagram of the SNO
detector~\cite{NIM}. The cavity was 34 meters high by 22 meters in
diameter at the equator, lined with a water- and radon-impermeable
Urylon plastic. The detector was situated 2 km underground in an
active nickel mine owned by Vale (formerly Inco Ltd) near Sudbury,
Ontario. The central element was 1000 tonnes of heavy water ($ >
99.5\% $ isotopically pure), on loan from AECL and housed in a
transparent acrylic vessel (AV) 12 meters in diameter and 5 cm thick.
The value of the heavy water was about \$300 million Canadian
dollars. The heavy water was viewed by 9438 20-cm diameter Hamamatsu
R1408 photomultiplier tubes (PMT) mounted on a stainless steel
geodesic photomultiplier support frame (PSUP). Each PMT had a 27-cm
entrance light concentrator to increase the effective photocathode
coverage to 54\%. A further 91 PMTs without concentrators were mounted
looking outward from the PSUP to observe events entering the detector
from the outside. The entire cavity outside the acrylic vessel was
filled with 7000 tonnes of ultra-pure ordinary water.

The construction sequence involved the building of the upper half of the
geodesic structure for the PMTs, installing them and lifting it with a
movable platform into place. This was followed by the construction of the
upper half of the acrylic vessel, which was a major process, involving the
bonding together of the first half of the 122 panels that were smaller than
the maximum length of 3.9 meters that could fit within the mine hoist cage.
The platform was then moved down by stages with the lower half of the acrylic
vessel and the PMT structure added sequentially.

Calibration was accomplished using a set of specialized sources that
could be placed on the central axis or on two orthogonal planes
off-axis in locations that covered more than 70\% of the
detector. These sources included 6.13-MeV $\gamma$~rays triggered from
decays of ${}^{16}{\rm{N}}$~\cite{bib:n16}, a source of
${}^8{\rm{Li}}$~\cite{bib:li8}, encapsulated sources of U, Th, a
${}^{252}{\rm{Cf}}$ fission neutron source, 19.8-MeV $\gamma$~rays from the
t(p,$\gamma$) reaction generated by a small accelerator suspended on
the central axis~\cite{bib:pt}. The ${}^{16}{\rm{N}}$ and
${}^8{\rm{Li}}$ were produced by a d(t,n) neutron source generated by
a small accelerator in a location near the SNO detector and
transported by capillary tubes to the main heavy water volume.

Signals from the SNO PMTs were received by electronics that made four
different measurements. For all PMT signals that were above a
threshold of the equivalent of 1/4 of a photoelectron of charge, the
electronics recorded a time relative to a global trigger, and provided
three different charge measurements: a short-window (60~ns)
integration of the PMT pulse, a long-window ($\sim$ 400~ns)
integration, and a low-gain version of the long-integration charge.
Each PMT above 1/4~pe also provided a 93 ns-wide analog trigger signal
and signals across the entire detector were summed together.  An event
was triggered if that sum exceeded a pre-set threshold, representing a
number of PMTs firing in coincidence. The system also kept absolute
time according to a GPS clock signal that was sent underground.

An accurate determination of the total solar neutrino flux required a
detector with ultra-low levels of any radioactive sources capable of
mimicking the signal. In addition, the residual levels needed to be
determined with sufficient accuracy that they contribute only slightly
to the overall measurement uncertainties.  Of particular concern for
SNO were two high-energy $\gamma$~rays produced in the
${}^{232}{\rm{Th}}$ and ${}^{238}{\rm{U}}$ chains (of energy 2615 and
2447~keV, respectively).  These were above the deuteron
photo-disintegration threshold and hence produce neutrons
indistinguishable from neutrino induced events.  As a consequence, all
the materials used in the fabrication of the detector were carefully
screened for radioactivity and the collaboration worked with
manufacturers to develop techniques to produce radioactively pure
materials and components.

To achieve this level of radiopurity in the water, both the light and
heavy water in SNO were purified through numerous stages including
filtration, degassing, customized ion-exchange and reverse osmosis.
The \hto\ and \dto\ purification plants were designed to remove Rn,
Ra, Th and Pb from the water, thereby eliminating sources giving rise
to the high energy $\gamma$~rays.  Two of the main elements of the SNO
\hto\ and \dto\ purification plants consisted of newly developed
ion-exchange processes using MnOx~\cite{bib:mnox} and
HTiO~\cite{bib:htio}, which targeted Ra, Th, and Pb nuclei in the
water.  With the removal of these elements, secular equilibrium was
broken and the short lived daughters quickly decayed away. The HTiO
and MnOx techniques developed by SNO were also used to assay the
amount of residual activity remaining in the fluids. In the case of
HTiO, the activity was eluted from HTiO by strong acids and
concentrated into liquid scintillator vials for counting. The
technique developed for MnOx used electrostatic counters to measure
the ${}^{222}{\rm{Rn}}$ and ${}^{220}{\rm{Rn}}$ emanating from the
surface.

Radon gas was particularly problematic as it emanated from materials
and could migrate or diffuse into sensitive areas of the
detector. Large process degassers and membrane contactors were used to
strip radon from the water with high efficiency. Monitoring degassers
were used to collect radon from the water into Lucas cells for a
determination of the residual contamination.

The design of the purification systems was to achieve a rate of
photo-disintegration events created by impurities of less than 10\% of
the NC rate predicted by the Standard Solar Model. To achieve this in
the \dto ~system required an equivalent of $< 3.8 \times 10^{-15}
{\rm{g Th}}/{\rm g}$\dto\ and $< 3.0 \times 10^{-14} {\rm{g U}}/{\rm g}$\dto.  The
requirements for the \hto\ outside the main detector were not as
stringent, and were $< 37 \times 10^{-15} {\rm{g Th}}/{\rm g}$\hto\ and $< 45
\times 10^{-14} {\rm{g U}}/{\rm g}$\hto. Measurements of the water purity
throughout the experiment showed that the levels for U in both \dto\ 
and \hto, and Th in \dto\ were consistently better than the design
value, while the Th content in \hto\ was about at the target
level. Hence the background contamination rate was not significant in
comparison to the neutrino NC signal.  The assay measurements were
consistent between HTiO, MnOx and radon gas measurements, and agreed
with {\it in-situ} measurements made with the PMT array.

\section{SNO Phase-I Physics Program}

SNO's first measurements of the rates of CC and NC reactions on
deuterium by $^8$B solar neutrinos used unadulterated D$_2$O in the
detector. The measurements had several challenges that differed from
the following two phases of the experiment. The first was that the
number of detected events expected from the NC reaction was low, in
part because the neutron capture cross section on deuterium is small,
but also because the energy of the $\gamma$~ray released in that
capture was just 6.25~MeV, near SNO's anticipated energy threshold.
The Phase-I data analysis was also the first to face unexpectedly
large instrumental backgrounds, which had to be removed before more
detailed analyses could proceed.  The primary result from Phase I was
a rejection of the null hypothesis that solar neutrinos do not change
flavor by comparing the flux measured by the CC reaction to those by
both NC and ES reactions.

In SNO Phase I, the signals from the ES, CC, and NC reactions could
not be separated on an event-by-event basis. Instead, a fit to the
data set was performed for each signal amplitude, using the fact that
they are distributed distinctly in the following three derived
quantities: the effective kinetic energy $T_{\rm eff}$ of the $\gamma$
ray resulting from the capture of a neutron produced by the NC
reaction or of the recoil electron from the CC or ES reactions, the
reconstructed radial position of the interaction ($R_{\rm fit}$) and
the reconstructed direction of the event relative to the expected
direction of a neutrino arriving from the Sun ($\cos
\theta_{\odot}$). The reconstructed radial positions $R_{\rm fit}$
were measured in units of AV radii and weighted by volume, so that
$\rho \equiv (R_{\rm fit}/R_{\rm AV})^3 = 1.0$ when an event reconstructs
at the edge of the \dto\ volume.

\begin{figure}[htb]
\begin{center}
\includegraphics[height=0.7\columnwidth]{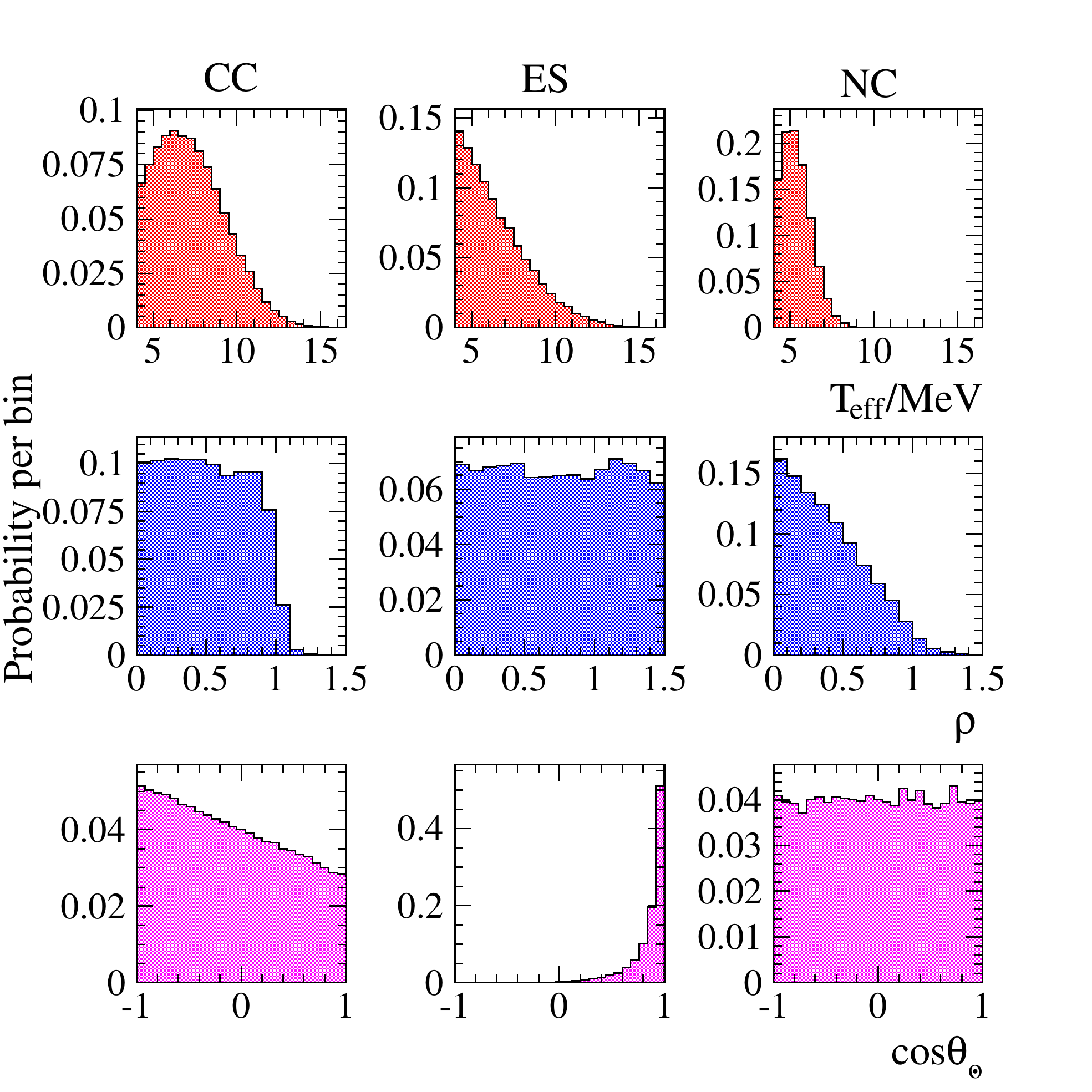}
\caption{The energy (top row), radial (middle row), and directional (bottom
row) distributions used to build PDFs to fit the SNO signal
data. $T_{\rm eff}$ is the effective kinetic energy of the $\gamma$
from neutron capture or of the electron from the ES or CC reactions, and
$\rho = (R_{\rm fit}/R_{\rm AV})^3$ is the reconstructed event radius,
volume-weighted to the 600~cm radius of the acrylic
vessel. \label{fig:pdfs}}
\end{center}
\end{figure}

Figure~\ref{fig:pdfs} shows the simulated distributions for each of
the signals.  The nine distributions were used as probability density
functions (PDFs) in a generalized maximum likelihood fit of the solar
neutrino data.  The top row shows the \teff\ distribution for each of
the three signals.  The CC and ES reactions both reflect the $^8$B
spectrum of incident neutrinos, with ES having a much softer spectrum
due to the kinematics of the reaction. The NC reaction is essentially
a line spectrum, because neutron capture on deuterium always results
in the same 6.25-MeV $\gamma$~ray.  The $\rho$ distributions are shown
in the middle row of Fig.~\ref{fig:pdfs}.  Electrons from the CC
reaction are distributed only within the heavy water volume, while
those from ES extend into the light water. The neutrons from the NC
reaction fall nearly linearly in $\rho$ from the center of the heavy
water to the edge, because of the probability of exiting the heavy
water volume and being captured on light water (and thus being below
the detection threshold).  The bottom row of Fig.~\ref{fig:pdfs} shows
the $\cos\theta_{\odot}$ distribution of the events.  The ES reaction
has a prominent peak indicating the solar origin for the
neutrinos. The CC electrons have a softer but nonetheless distinctive
$\sim$ $(1-1/3\cos\theta_{\odot})$ distribution, while the NC neutrons
have no correlation at all with the solar direction.

The Phase-I data set was acquired between November 2, 1999 and May 31,
2001, and represented a total of 306.4 live days.  The SNO detector
responded to several triggers, the primary one being a coincidence of
18 or more PMTs firing within a period of 93~ns (the threshold was
lowered to 16 or more PMTs after December 20, 2000). The rate of such
triggers averaged roughly 5~Hz.  A ``random'' trigger also pulsed the
detector at 5~Hz throughout the data acquisition period.

To provide a final check against statistical bias, the data set was
divided in two: an ``open'' data set to which all analysis procedures
and methods were applied, and a ``blind'' data set upon which no
analysis within the signal region (between 40 and 200 hit PMTs) was
performed until the full analysis program had been finalized.  The
blind data set began at the end of June 2000, at which point only 10\%
of the data set was being analyzed, leaving the remaining 90\% blind.
The total size of the blind data set thus corresponded to roughly 30\%
of the total live time.

The presence of many sources of events created by the instrumentation
of the SNO detector was apparent even before the start of heavy water
running.  The approach to removing these events began with a suite of
simple cuts to act as a series of ``coarse filters,'' removing the
most obvious of such events, before any event reconstruction.  Sources
of instrumental events included light generated by the PMTs (``flasher
PMTs'') that happened for every PMT and occurred at a rate of roughly
1/minute; light from occasional high-voltage breakdown in the PMT
connector or base; light generated by static discharge in the neck of
the vessel; electronic pickup; and isotropic light occasionally
emitted by the acrylic vessel.  The cuts were based only on simple
low-level information such as PMT charges and times, but the full
suite removed the vast majority of the instrumental events.  Two
independent suites were created to help validate the overall
performance of the coarse filters.  The acceptance for signal events
of the instrumental background cuts was measured using calibration
source data, and was found to be $>$99.5\%.

The reconstruction of event position, direction, and energy was performed
on events that passed the instrumental background cuts.  Position reconstruction
used the relative PMT-hit times as well as the angular distribution of
photon hits about a hypothesized event direction.  Event energy
used the number of PMT hits along with an analytic model of the detector
response to Cherenkov light that was a function of event position and
direction. For both position and energy, additional
independent algorithms were used to validate the
results~\cite{longd2o}.

After reconstruction, a further set of cuts were applied to remove
events that were not consistent with the timing and angular distribution of
Cherenkov light (``Cherenkov Box Cuts'').   The two cuts that defined the
Cherenkov Box were the width of the prompt timing peak of the PMT hits, and the
average angle between pairs of hit PMTs.

Neutrons and events from spallation products that were created by the
passage of muons or the interactions of atmospheric neutrinos were
removed by imposing a 20-s veto window following the muon events, and
a 250-ms veto following any event that produced more than 60 fired
PMTs (roughly 7 MeV of electron-equivalent total energy $E_{\rm eff}$).
The final set of cuts were the requirement that events have a
reconstructed effective kinetic energy
$T_{\rm eff}=E_{\rm eff}-0.511~{\rm MeV}>5.0$~MeV, 
and a reconstructed position with $R_{\rm fit}<550$~cm ($\rho<0.77$).

For SNO Phase I to be able to make a measurement of the total flux of
neutrinos via the NC reaction, it was critical that the number of
background neutrons was small compared to those expected from solar
neutrinos.  The most dangerous source of such neutrons was those from
the photodisintegration of deuterons by $\gamma$~rays, resulting from
decays in the $^{238}$U and $^{232}$Th chains.  The levels of U and Th
were measured in two ways: {\it ex situ} assays of the heavy and light
water~\cite{bib:mnox,bib:htio}, and {\it in situ} measurements of
$^{208}$Tl and $^{214}$Bi concentrations using the differences in the
isotropy of their Cherenkov-light distributions. Both methods agreed
well, and by combining them the levels of U and Th in the heavy water
were found to be:
\begin{eqnarray} 
^{232}\mbox{Th} &:& 1.61 \pm 0.58\times 10^{-15} \mbox{g Th/g D$_2$O} \nonumber \\ 
^{238}\mbox{U} &:& 17.8 ^{+3.5}_{-4.3} \times 10^{-15} \mbox{ g U/g D$_2$O}. \nonumber 
\end{eqnarray}

With these measurements, and those of radioactivity in the light water
and acrylic vessel, the total number of background neutrons from
photodisintegration in the Phase-I data set was 38.2$^{+9.4}_{-9.5}$
from the $^{232}$Th chain and 33.1$^{+6.7}_{-7.1}$ from $^{238}$U
chain. Neutrons from other sources, such as atmospheric neutrinos and
$(\alpha,n)$ processes, were found to be just $7^{+3}_{-1}$~counts.

The PDFs shown in Fig.~\ref{fig:pdfs} were created via a calibrated
and over-constrained Monte Carlo simulation.  Events resulting from
$^8$B neutrino interactions or sources of background were passed
through a detector model that included the propagation of electrons,
$\gamma$~rays, and neutrons through the heavy water, a detailed
optical response of the detector media and PMTs, and data acquisition
electronics.  Parameters such as optical attenuation lengths,
scattering, and overall PMT collection efficiency were measured by
deploying a diffuse laser source~\cite{laserball} and a $^{16}$N
source~\cite{bib:n16} of 6.1-MeV $\gamma$~rays throughout the detector
volume.  Residual differences between the model prediction for energy
scale, energy resolution, vertex reconstruction bias and vertex
resolution, were taken as systematic uncertainties on the model, and
were within $\pm$ 1\%.  The overall neutron capture efficiency was
measured using the deployment of a $^{252}$Cf source throughout the
detector volume. 

The fit to the data set using the PDFs of Fig.~\ref{fig:pdfs} was done
via an extended log-likelihood of the form:
\begin{equation}
\log L = -\sum_i N_i + \sum_j n_j \ln\{\nu(T_{{\rm eff} j},\rho_j, \cos \theta_{\odot j})\},
\end{equation}
where $N_i$ is the number of events of type $i$ (e.g. CC, ES, or NC),
$j$ is a sum over all three-dimensional bins in the three signal
extraction parameters \teff, $\rho$, and $\cos \theta_\odot$, and
$n_j$ is the number of detected events in each bin.  The numbers of
CC, ES, and NC events were treated as free parameters in the fit.  The
likelihood function was maximized over the free parameters, and the
best fit point yielded the number of CC, ES, and NC events along with
a covariance matrix.

The Phase-I data set was fit under two different assumptions. The
first was that the recoil electron spectra of the CC and ES events
resulted from an undistorted $^8$B neutrino spectrum, thus testing the
null hypothesis that solar neutrinos do not change flavor. The second
fit had no such constraint, and could be done either by fitting events
bin-by-bin in energy~\cite{snocc} or by using only $\rho$ and $\cos
\theta_\odot$~\cite{snonc}.

In addition to fitting for the three signal rates (CC, ES, and NC),
the SNO data also allowed a direct fit for the neutrino flavor content
through a change of variables:
\begin{eqnarray}
\label{eq.flavor_fit}
\phi_{\rm CC} & = & \phi(\nu_e) \\
\phi_{\rm ES} & = & \phi(\nu_e) + 0.1559 \phi(\nu_{\mu\tau}) \\
\phi_{\rm NC} & = & \phi(\nu_e) + \phi(\nu_{\mu\tau}). 
\end{eqnarray}
The factor of 0.1559 is the ratio of the ES cross sections for
$\nu_{\mu\tau}$ and $\nu_e$ above $T_{\rm eff}=5.0$~MeV.  Making this
change of variables and fitting directly for the flavor content, the
null hypothesis test of no flavor change is reduced to a test of
$\phi(\nu_{\mu\tau})=0$.

Conversion of event numbers from the fit into neutrino fluxes required
corrections for cut acceptance, live time, measured neutron capture
efficiency, subtraction of neutron backgrounds, and effects not
included in the Monte Carlo simulation (such as the eccentricity of
the Earth's orbit). With these corrections applied, and measurements
of the systematic uncertainties on both acceptances and detector
response, the flux values for the constrained fit are~(in units of
$10^6~{\rm cm}^{-2} {\rm s}^{-1}$):
\begin{center}
$\phi_{\rm CC} =
  1.76^{+0.06}_{-0.05}\mbox{(stat.)}^{+0.09}_{-0.09}~\mbox{(syst.)}$ \\
$\phi_{\rm ES} = 2.39^{+0.24}_{-0.23}\mbox{(stat.)}^{+0.12}_{-0.12}~\mbox{(syst.)}$ \\
$\phi_{\rm NC} = 5.09^{+0.44}_{-0.43}\mbox{(stat.)}^{+0.46}_{-0.43}~\mbox{(syst.)}.$ \\
\end{center}

The physical interpretation of the ``flux'' for each interaction type
is that it is the equivalent flux of $^8$B $\nu_e$s produced from an
undistorted energy spectrum that would yield the same number of
events inside the signal region from that interaction as was seen in
the data set. 

\begin{figure}[htb]
\begin{center}
\includegraphics[width=0.65\columnwidth]{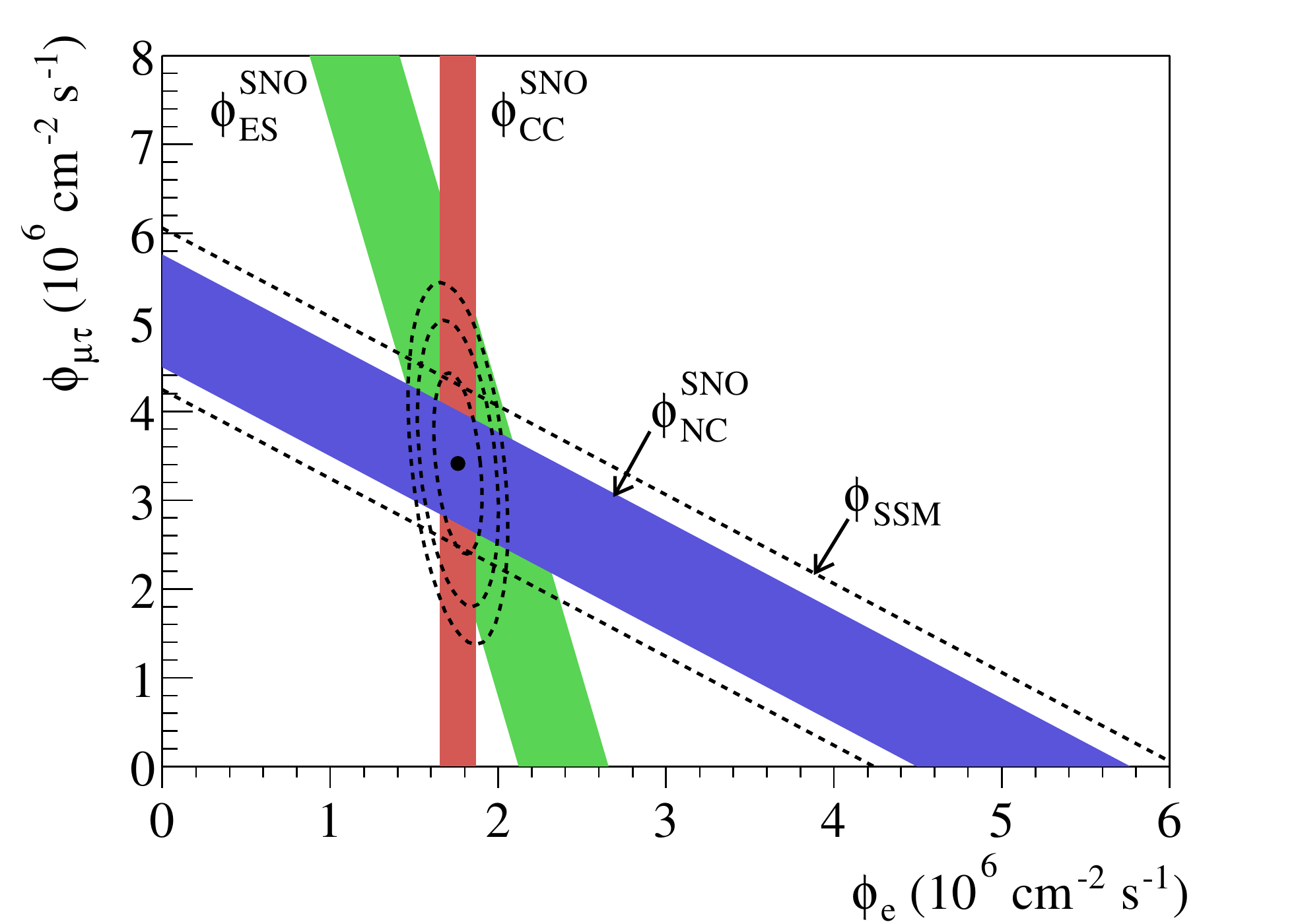}
\caption{\label{fig:hime_plot}Flux of ${}^{8}$B solar neutrinos which are
$\mu$ or $\tau$ flavor vs flux of electron neutrinos deduced from the
three neutrino reactions in SNO.  The diagonal bands show the total
${}^{8}$B flux as predicted by the BP2000 SSM~\cite{bp2000} (dashed lines)
and that measured with the NC reaction in SNO (solid band).  The
intercepts of these bands with the axes represent the $\pm 1\sigma$
errors.  The bands intersect at the fit values for $\phi_{e}$ and
$\phi_{\mu\tau}$, indicating that the combined flux results are
consistent with neutrino flavor transformation with no distortion
in the ${}^{8}$B neutrino energy spectrum.}
\end{center}
\end{figure}

The inequality of the fluxes determined from the CC, ES, and NC
reactions provided strong evidence for a non-$\nu_e$ component to the
$^8$B solar neutrinos.  Figure~\ref{fig:hime_plot} shows the
constraints on the flux of $\nu_e$ versus the combined $\nu_{\mu}$ and
$\nu_\tau$ fluxes derived from the CC, ES, and NC rates.  Together the
three rates were inconsistent with the hypothesis that the $^8$B flux
consists solely of $\nu_e$s, but are consistent with an admixture
consisting of about $1/3~\nu_e$ and $2/3~ \nu_\mu$ and/or $\nu_\tau$.

Changing variables to provide a direct measure of flavor content, the 
fluxes are~(in units of $10^6~{\rm cm}^{-2} {\rm s}^{-1}$):
\begin{center}
$\phi(\nu_e) =
  1.76^{+0.05}_{-0.05}\mbox{(stat.)}^{+0.09}_{-0.09}~\mbox{(syst.)}$ \\
$\phi(\nu_{\mu\tau}) =
  3.41^{+0.45}_{-0.45}\mbox{(stat.)}^{+0.48}_{-0.45}~\mbox{(syst.)}. $\\
\end{center}
Adding the statistical and systematic errors in quadrature,
$\phi(\nu_{\mu\tau})$ is $5.3\sigma$ away from its null hypothesis
value of zero.

With the corrections applied and normalizing to the Monte Carlo event
rates, the ``NC flux'' for the energy-unconstrained fit between 
$5 < T_{\rm eff} < 19.5$~MeV (using only $\rho$ and $\cos \theta_{\odot}$), was:
\begin{center}
$\phi_{\rm NC} =
  6.42^{+1.57}_{-1.57}\mbox{(stat.)}^{+0.55}_{-0.58}~\mbox{(syst.)} \times
10^{6}$~cm$^{-2}$~s$^{-1}$. \\
\end{center}

Both measurements of the total active fluxes $\phi_{NC}$, as well as
the sum of $\phi(\nu_e)+\phi(\nu_{\mu\tau})$, were in good agreement
with Standard Solar Model predictions\cite{bp2000,TC}.  Using the same
data set, SNO did not observe any statistically significant day-night
asymmetries of the CC, NC, and ES reaction
rates~\cite{bib:d2odaynight}.

These results for the full data set of Phase I were in good agreement
with and more accurate than the results obtained~\cite{snocc} by
comparison of the SNO CC data with ES data from Superkamiokande.

\section{SNO Phase-II Physics Program}
\label{sec:sno_ii}

In Phase II, approximately 2000~kg of NaCl was dissolved in the
1000-tonne heavy-water neutrino target of SNO.  The addition of salt
enhanced the experiment's sensitivity to detect $^8$B solar neutrinos
through the NC reaction in several ways.  The thermal neutron capture
cross section for $^{35}$Cl is nearly five orders of magnitude larger
than that for the deuteron, resulting in a significant increase in the
neutron capture efficiency in the detector.  The $Q$-value for
radiative neutron capture on $^{35}$Cl is 8.6~MeV, which is 2.3~MeV
above that for capture on the deuteron.  The increase in the released
energy led to more observable NC events above the energy threshold
($\teff >$~5.5~MeV) in the measurement, but more importantly, the
cascade of prompt $\gamma$~rays following neutron capture on $^{35}$Cl
produced a Cherenkov-light hit pattern on the PMT array that was
significantly different from that produced by a single relativistic
electron from the CC or the ES reactions.  Multiple $\gamma$~rays
produced a more isotropic pattern of triggered PMTs on the PSUP.  This
difference in the observed event topology allowed the statistical
separation between events from the NC and the CC reactions without
making any assumption on the underlying neutrino energy spectrum.

The complete Phase-II data set consisted of $391.432 \pm 0.082$ live
days of data recorded between July 26, 2001 and August 28, 2003.  A
blind analysis was performed on the initial 254.2-live-day data set in
Ref.~\cite{saltprl}, followed by an analysis of the full data set in
Ref.~\cite{nsp}.  In the blind analysis, an unknown fraction of the
data were excluded, and an unknown admixture of neutrons following
cosmic muons events was added.  An unknown scaling factor of the NC
cross section was also applied to the simulation code.  After fixing
all analysis procedures and parameters, the blindness constraints were
removed for a full analysis of the 254-live-day data set.

\begin{figure}
\begin{center}
\includegraphics[width=0.6\textwidth]{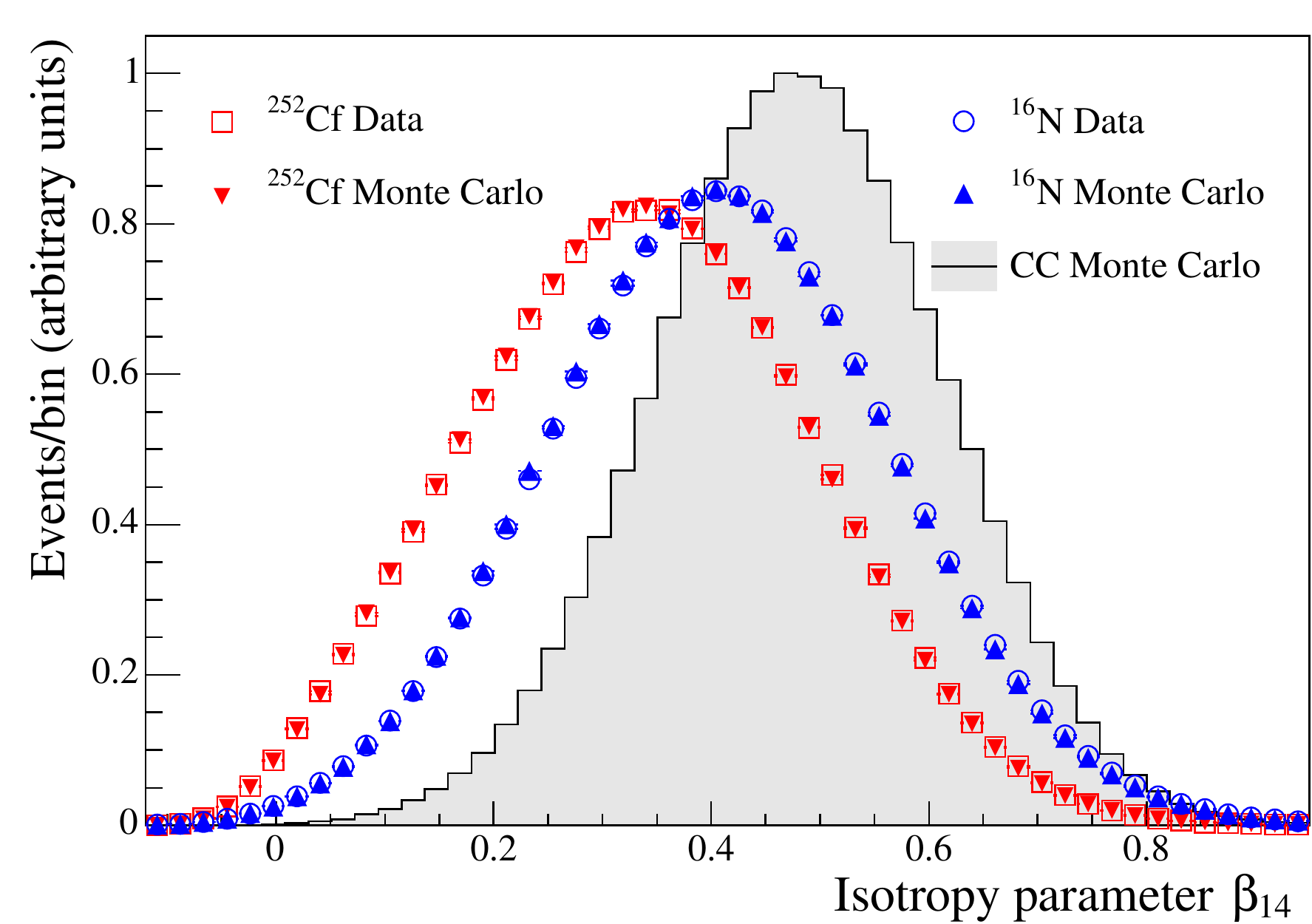}
\caption{\label{fig:prl_isotropy} $\beta_{14}$ isotropy distributions
for \cf\ data and MC, \ns\ data and simulations, and simulated CC 
events.  Good agreement was found between simulated $\beta_{14}$ and 
\cf\ and \ns\ calibration data.  Note that the distribution 
normalizations are arbitrary and chosen to allow the shape 
differences to be seen clearly.}
\end{center}
\end{figure}

To exploit the difference in Cherenkov-light event topology for
different types of signals, several variables were constructed.  The
variable that was eventually adopted, which could be simply
parameterized and facilitated systematic uncertainty evaluations, was
$\beta_{14}\equiv\beta_1+4\beta_4$ where
\begin{equation}
\beta_l = \frac{2}{N(N-1)}\sum_{i=1}^{N-1} \sum_{j=i+1}^N
P_l(\cos\theta_{ij}).
\end{equation}
In this expression $P_l$ is the Legendre polynomial of order $l$,
$\theta_{ij}$ is the angle between triggered PMTs $i$ and $j$ relative
to the reconstructed event vertex, and $N$ is the total number of
triggered PMTs in the event.  Figure~\ref{fig:prl_isotropy} shows the
difference in the $\beta_{14}$ distributions between neutron (NC) and
electron (CC or ES) events.

The neutron response of the detector was calibrated primarily with
neutrons produced by a \cf\ source with secondary checks made by
analysis of neutrons generated by an Am-Be source and by Monte Carlo
simulations.  The volume-weighted detection efficiency for neutrons
generated uniformly in the \dto\ for the analysis threshold of
$\teff=5.5$~MeV and a fiducial volume of 550~cm ($\rho<0.77$) was
found to be 0.407$\pm$0.005~(stat.)$^{+0.009}_{-0.008}$~(syst.).

As in Phase I, a normalization for photon detection efficiency based
on \nsixtn~\cite{bib:n16} calibration data and Monte Carlo simulations
was used to set the absolute energy scale.  A $\sim$2\% gain drift was
observed in the \nsixtn\ data taken throughout the running period;
this drift was predicted by simulations based on temporal changes in
the optical measurements.  The overall energy-scale resolution
uncertainty was found to be 1.15\%.

Compared to Phase I, the addition of salt increased the sensitivity to
neutron capture at large $\rho$, making it possible to detect
background neutrons originating at or near the acrylic vessel and in
the \hto.  In Phase I, the magnitude of these ``external source''
neutrons were estimated and fixed in the neutrino signal decomposition
analysis.  In Phase II, the amplitude of the $\rho$ PDF of the
external source neutrons was allowed to vary in the maximum likelihood
fit.

\begin{figure}
\begin{center}
\includegraphics[width=0.45\textwidth]{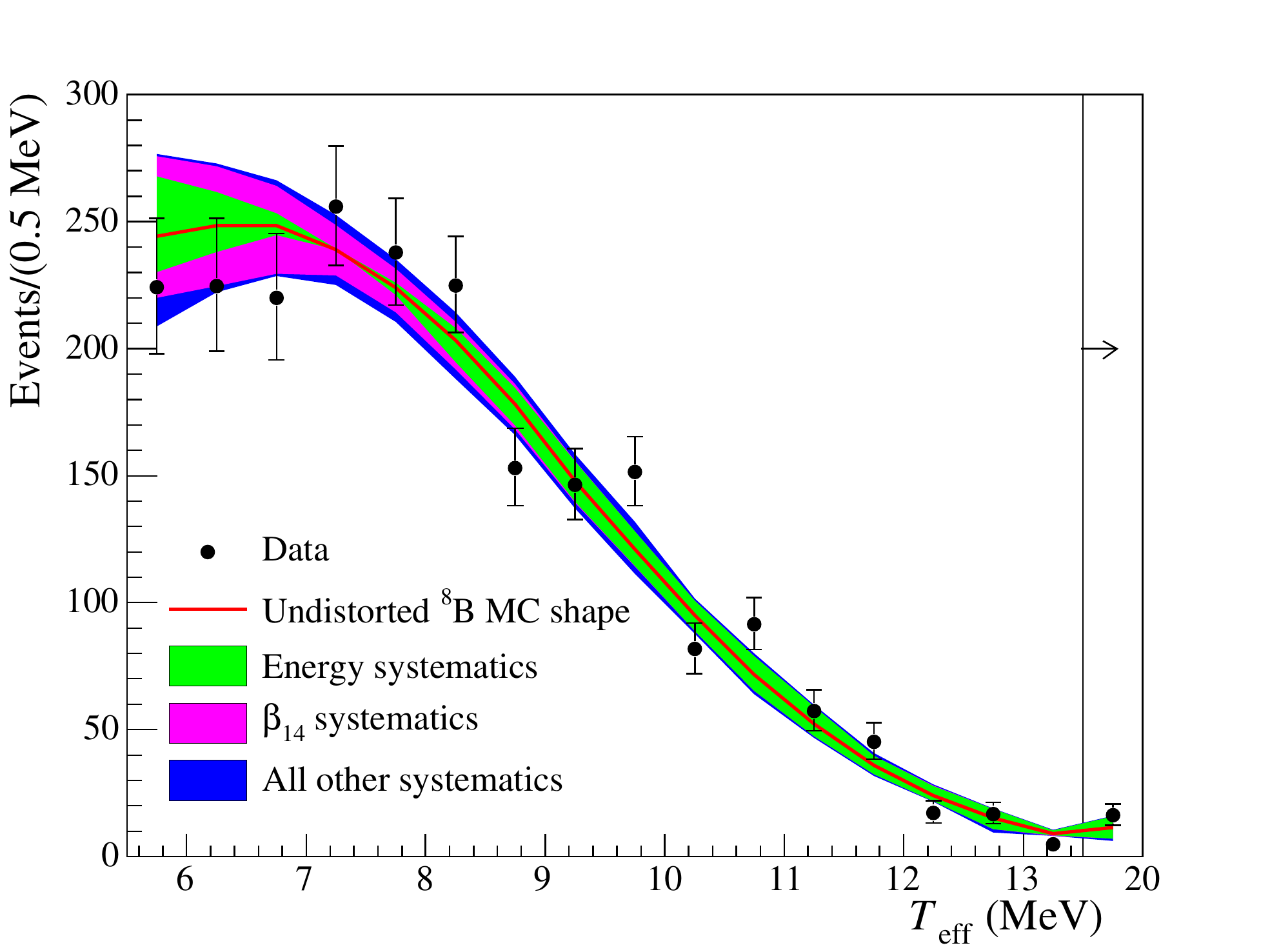}
\includegraphics[width=0.45\textwidth]{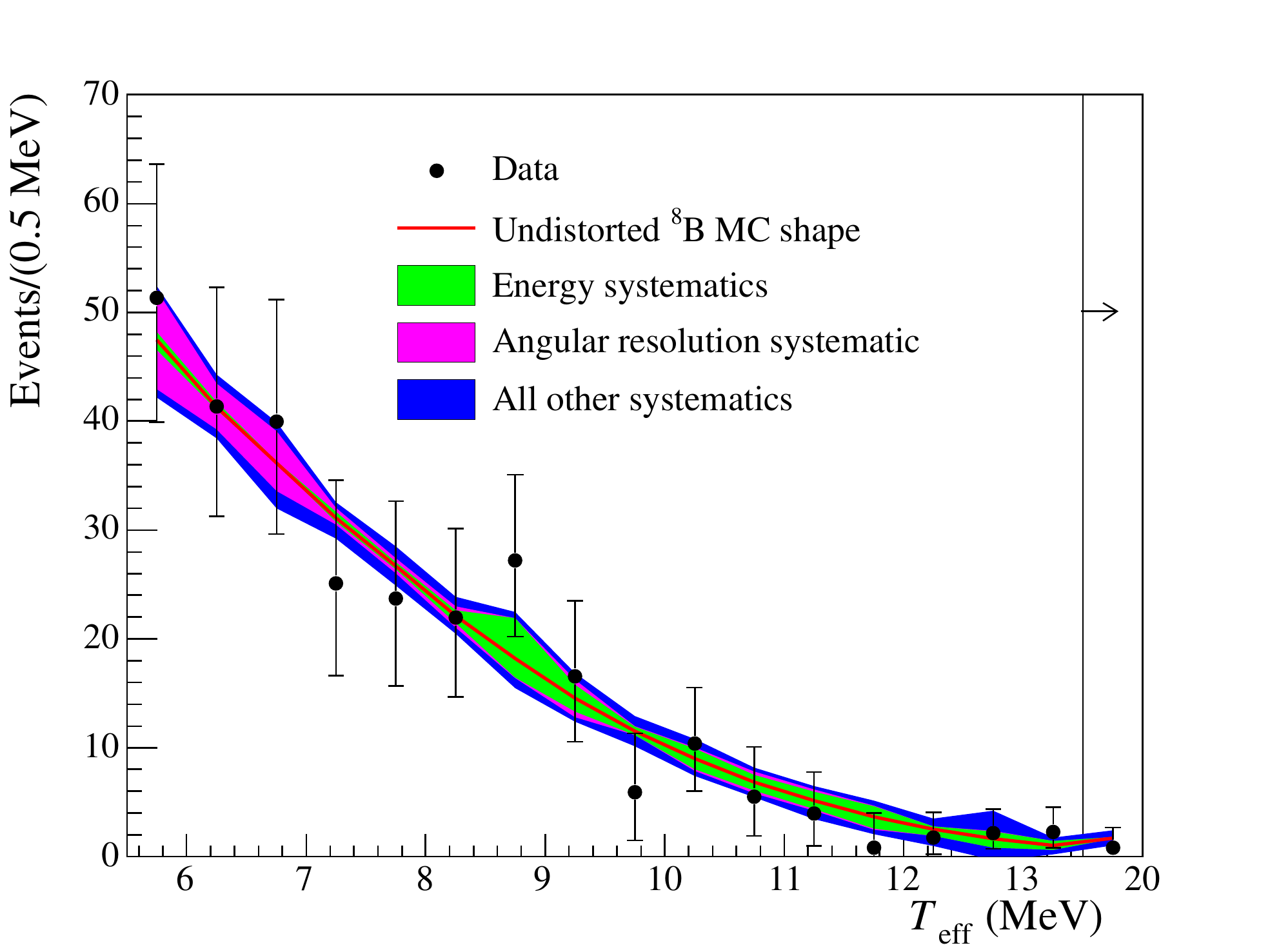}
\caption{\label{fig:cc_es_fit}{Left: Extracted CC $\teff$ spectrum 
with statistical error bars compared to predictions for an 
undistorted $^8$B shape with combined systematic uncertainties, 
including both shape and acceptance components.  The 
highest-energy bin represents the average number of events per 
0.5~MeV for the range of 13.5-20~MeV.  Right: An analogous plot 
for the extracted ES $\teff$ spectrum.}}
\end{center}
\end{figure}

In the determination of the electron-energy spectrum from CC and ES
interactions and the total active solar neutrino flux, an extended
maximum likelihood fit with four data variables (\teff, $\rho$,
$\cos\theta_\odot$, and $\beta_{14}$) was performed. To obtain the
electron energy spectra of CC and ES interactions, probability density
functions (PDFs) were simulated for $\teff$ intervals, which spanned
the range from $5.5$~MeV to $13.5$~MeV in $0.5$~MeV steps.  A single
bin was used for $\teff$ values between $13.5$ and $20$~MeV.  The
\teff\ PDFs for NC and external source neutrons were simply the
detector's energy response to radiative neutron captures on $^{35}$Cl
and $^2$H.  Minor adjustments were applied to the PDFs to take into
account signal loss due to instrumental cuts not modeled by the
simulation.  A four-dimensional PDF was implemented in the signal
decomposition:
\begin{equation}
 P(\teff,\beta_{14},\rho, \cos\theta_\odot) = P(\teff,\beta_{14},\rho) \times
P(\cos\theta_\odot | \teff,\rho),
\label{eq:threed}
\end{equation}
where the first factor is just the 3-dimensional PDF for the variables
$\teff$, $\beta_{14}$, and $\rho$, while the second factor is the
conditional PDF for $\cos\theta_\odot$, given $\teff$ and $\rho$.  In
the maximum likelihood fit the PDF normalizations for CC and ES
components were allowed to vary separately in each $\teff$ bin to
obtain their model-independent spectra. For the NC and external
neutron components only their overall normalizations were allowed to
vary.  Figure~\ref{fig:cc_es_fit} shows the extracted CC and ES
electron energy spectra.

For this energy-unconstrained analysis, the integral neutrino flux
were determined to be~(in units of $10^6~{\rm cm}^{-2} {\rm s}^{-1}$):
\begin{eqnarray*}
\phi^{\text{uncon}}_{\text{CC}} & = & 1.68^{+0.06}_{-0.06}\mbox{(stat.)}^{+0.08}_{-0.09}\mbox{(syst.)} \\
\phi^{\text{uncon}}_{\text{ES}} & = & 2.35^{+0.22}_{-0.22}\mbox{(stat.)}^{+0.15}_{-0.15}\mbox{(syst.)} \\
\phi^{\text{uncon}}_{\text{NC}} & = & 4.94^{+0.21}_{-0.21}\mbox{(stat.)}^{+0.38}_{-0.34}\mbox{(syst.)}~\mbox{,}
\end{eqnarray*} 
and the ratios of the CC flux to that of NC and ES are
\begin{eqnarray*}
\frac{\phi^{\text{uncon}}_{\text{CC}}}{\phi^{\text{uncon}}_{\text{NC}}}
= 0.340\pm 0.023~\mbox{(stat.)}~^{+0.029}_{-0.031}~\mbox{(syst.)} \\
\frac{\phi^{\text{uncon}}_{\text{CC}}}{\phi^{\text{uncon}}_{\text{ES}}}
= 0.712 \pm 0.075~\mbox{(stat.)~}^{+0.045}_{-0.044}~\mbox{(syst.)}. \\
\end{eqnarray*}

In a subsequent analysis of the combined Phase-I and Phase-II data
sets~\cite{bib:leta}, the energy threshold was lowered to
$\teff>$3.5~MeV (the lowest achieved with a water Cherenkov neutrino
detector).  Two different analysis methods, one based on binned
histograms and another on kernel estimation, were developed in the
joint analysis.  With numerous improvements to background modeling,
optical and energy response determination, and treatment of systematic
uncertainties in the signal decomposition process, the uncertainty in
the total active solar neutrino flux was reduced by more than a factor
of two~(in units of $10^6~{\rm cm}^{-2} {\rm s}^{-1}$):
\begin{equation}
\phi^{\text{uncon}}_{\text{NC}}  =  5.140^{+0.160}_{-0.158}\mbox{(stat.)}^{+0.132}_{-0.117}\mbox{(syst.)}.
\end{equation}
If the unitarity condition is assumed (i.e. no transformation from
active to sterile neutrinos), the CC, ES and NC rates are directly
related to the total $^8$B solar neutrino flux.  A signal
decomposition fit was performed in this combined analysis in which the
free parameters directly described the total $^8$B neutrino flux and
the energy-dependent $\nu_e$ survival probability.  In this scenario,
the total $^8$B neutrino flux was found to be~(in units of 
$10^6~{\rm cm}^{-2} {\rm s}^{-1}$):
\begin{equation}
 \Phi_{^{8}\text{B}}  =  5.046^{+0.159}_{-0.152}\mbox{(stat.)}^{+0.107}_{-0.123}\mbox{(syst.)}.
\end{equation}
Further details on this joint analysis and that for data from all three phases of the experiment can be found in Sec.~\ref{sec:combined}.

\section{SNO Phase-III Physics Program}
\label{sec:sno_iii}

In Phase III of the experiment, an array of $^3$He proportional
counters~\cite{ncdcountersnim} was deployed in the \dto\ volume.  The
neutron signal in the inclusive total active neutrino flux measurement
was detected predominantly by this ``Neutral-Current Detection'' (NCD)
array via
\begin{displaymath} n + \mbox{$^3$He} \rightarrow p + t + 764\mbox{ keV},
\end{displaymath} and was separate from the Cherenkov-light signals in the
$\nu_e$ flux measurement.  The separation resulted in reduced
correlations between the total active neutrino flux and $\nu_e$ flux
measurements, and therefore the measurement of the total active $^8$B
solar neutrino flux was largely independent of the methods of previous
phases.

\begin{figure}[htb!]
\begin{center}
\includegraphics[width=0.4\textwidth]{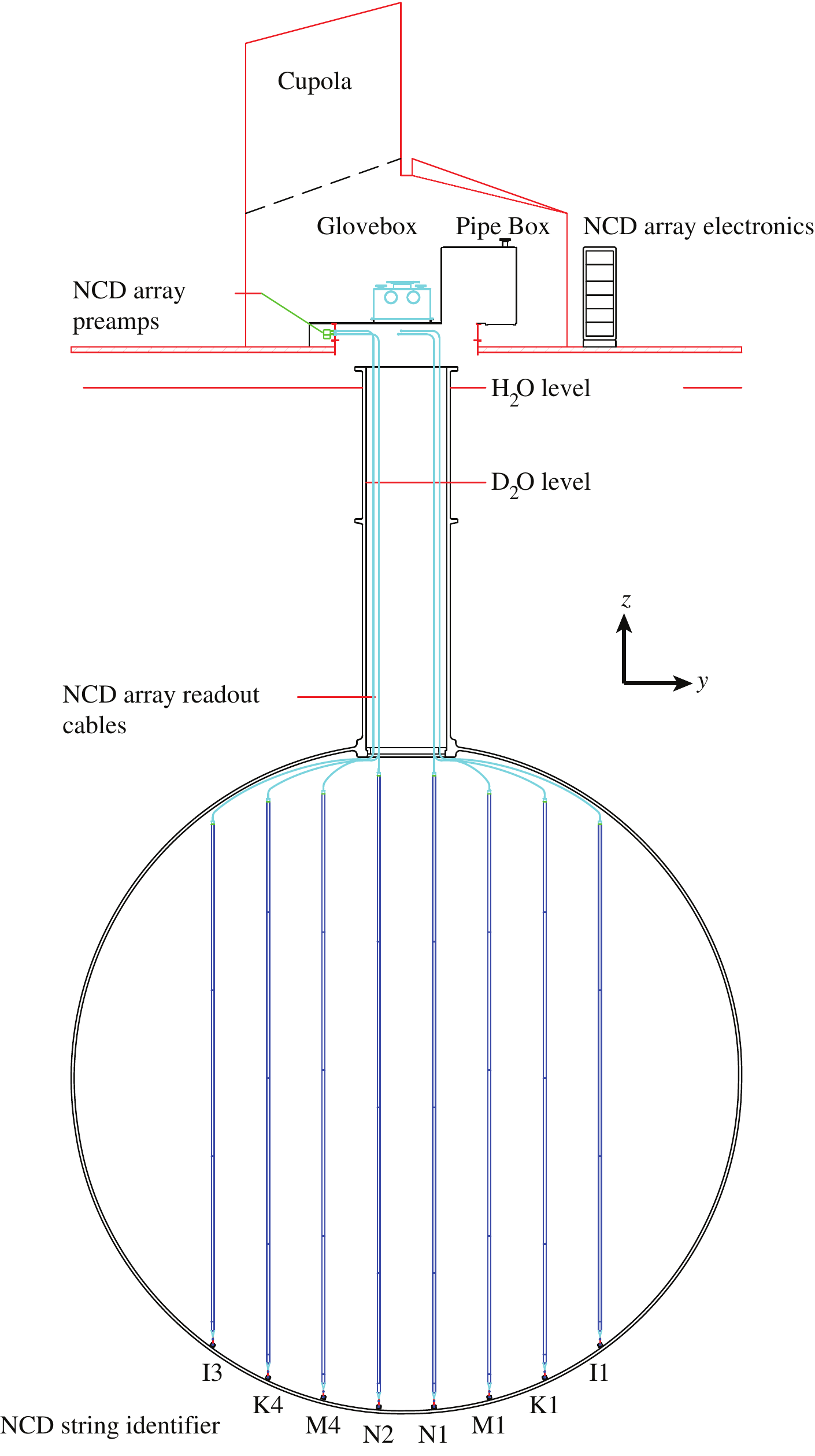}
\caption{Side view of the SNO detector in Phase III.  Only the first 
row of NCD strings from the $y-z$ plane are displayed in this 
figure.\label{fig:NCD_ZYview} }
\end{center}
\end{figure}

The NCD array consisted of 36 strings of $^3$He and 4 strings of
$^4$He proportional counters, which were deployed on a square grid
with 1-m spacing~\cite{ncdcountersnim}.  The $^4$He strings were not
sensitive to neutrons and were used for characterizing non-neutron
backgrounds.  Each detector string was made up of three or four
individual 5-cm-diameter counters that were laser-welded together.
The counters were constructed from ultra-pure nickel produced by a
chemical deposition process to minimize internal radioactivity.
Figure~\ref{fig:NCD_ZYview} shows a side view of the SNO detector with
the NCD array in place. 
 
The Phase-III data set represented $385.17 \pm 0.14$~live days of data
recorded between November 27, 2004 and November 28, 2006.  During this
period, the SNO detector was live nearly 90\% of the time, with
approximately 30\% of the live time spent on detector calibration.
Six $^3$He strings were defective and their data were excluded in the
measurement.

In Phase III, optical and energy calibration procedures, as well as 
Cherenkov-event reconstruction, were modified from those in previous
phases to account for the optical complexity introduced by the NCD array.
Similar to previous phases, the primary source for energy scale and
resolution calibration of the PMT array was the $^{16}$N
source~\cite{bib:n16}.  In Phase III, the energy scale uncertainty was
found to be 1.04\%.

The NCD array had two independently triggered readout systems, a fast
shaper system that recorded signal peak heights and could operate at
high rates in the event of a galactic supernova, and a slower, full
waveform digitization system that had a 15-$\mu$s window around the
signal. The detector signal response to neutrons was calibrated using
Am-Be neutron source data.

The principal method for determining the neutron detection efficiency
of the PMT and NCD arrays was to deploy an evenly distributed
$^{24}$Na source in the \dto~\cite{na24_nim_paper}. The source was
deployed by injecting a neutron-activated brine throughout the volume.
The $\gamma$s created by the $^{24}$Na then created free neutrons
through photodisintegration of deuterons in the heavy water.  Thus
neutron capture efficiency determined this way was found to be
$\epsilon=0.211\pm 0.005$. Additional corrections for threshold and
other effects reduced the overall detection efficiency to 86.2\% of
this value.

A small fraction of NC neutrons was captured by the deuterons in the target,
resulting in the emission of a 6.25-MeV $\gamma$~ray that could be detected by
the PMT array.  The efficiency for the detection of these events was
$0.0502\pm0.0014$.

The evaluation of
the intrinsic radioactive backgrounds in the detector construction materials
and in the \dto\ and \hto\ volumes followed analogous procedures in previous
phases, with adjustments for the added optical complexity of the detector, and
with new analyses developed to measure backgrounds on the NCDs
themselves.  These analyses used both information from Cherenkov light and
signals from the NCD counters, and the two techniques were in good
agreement.  Two radioactive ``hot spots'' were identified on two separate NCD strings from
the Cherenkov-light signals.  An extensive experimental program was developed
to measure the radioactive content of these hot spots.  More details can be
obtained from Ref.~\cite{HSNIM}.

Like Phases I and II, extraction of the neutrino signals  
for Phase III used an extended maximum likelihood fit to data, which for
this phase included both PMT (Cherenkov) signals and the summed energy spectrum
from the NCD shaper data (``shaper energy'', $E_{\textrm{NCD}}$). 
The fit to the shaper energy included an alpha background
distribution~\cite{ncdsim} from simulation, a neutron spectrum determined 
from $^{24}$Na calibration source data, expected neutron backgrounds, 
and instrumental background event
distributions.  The same blindness approach was used here as in Phase II.

\begin{figure}
\begin{center}
\includegraphics[width=0.60\columnwidth]{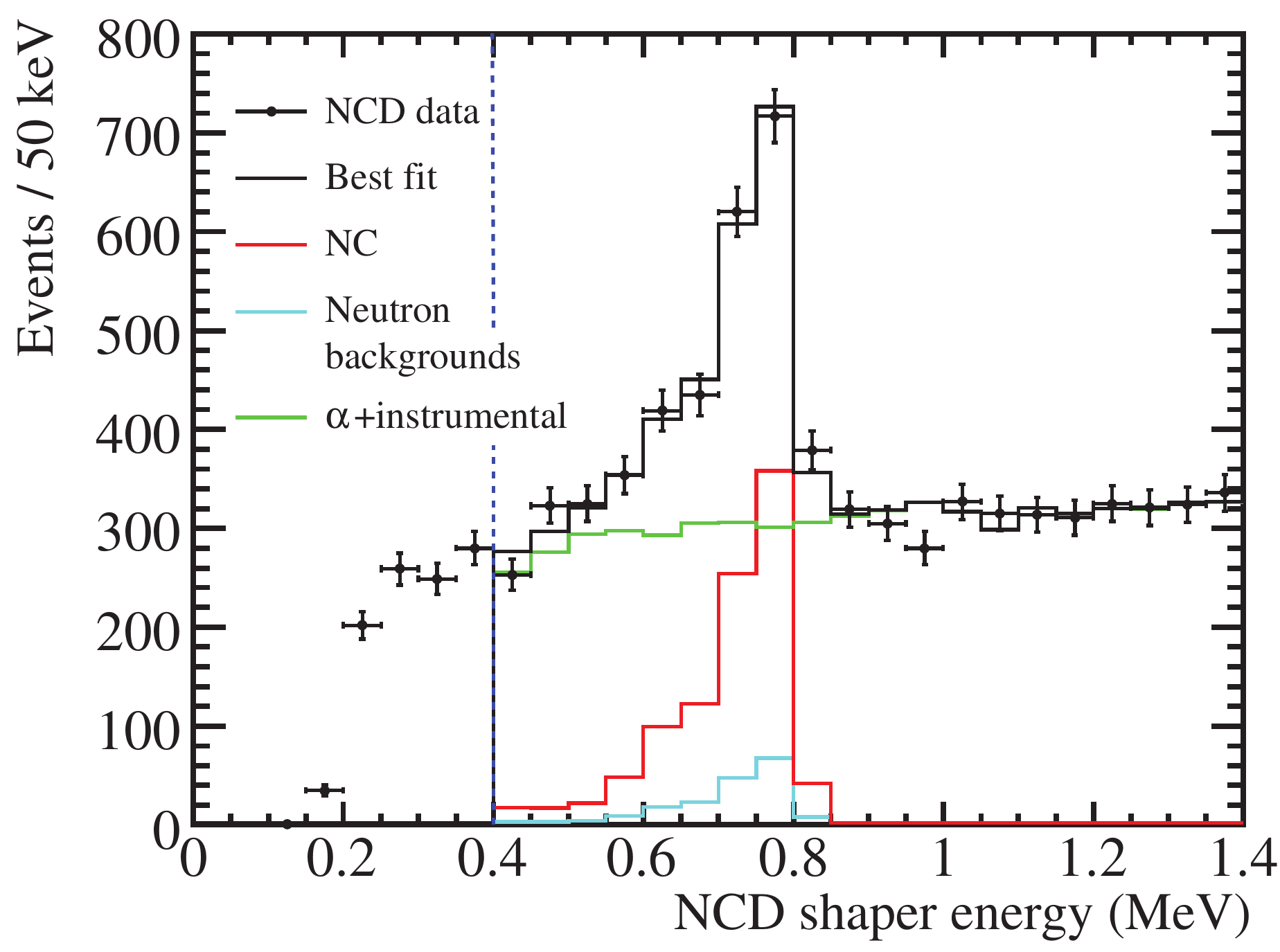}
\caption{\label{fig:ncdshaper} One-dimensional projection of NCD-array 
shaper-energy data overlaid with best-fit results for the NC signal as 
well as for the neutron, alpha and instrumental backgrounds.}  
\end{center}
\end{figure}

\begin{figure}
\begin{center}
\includegraphics[width=0.60\columnwidth]{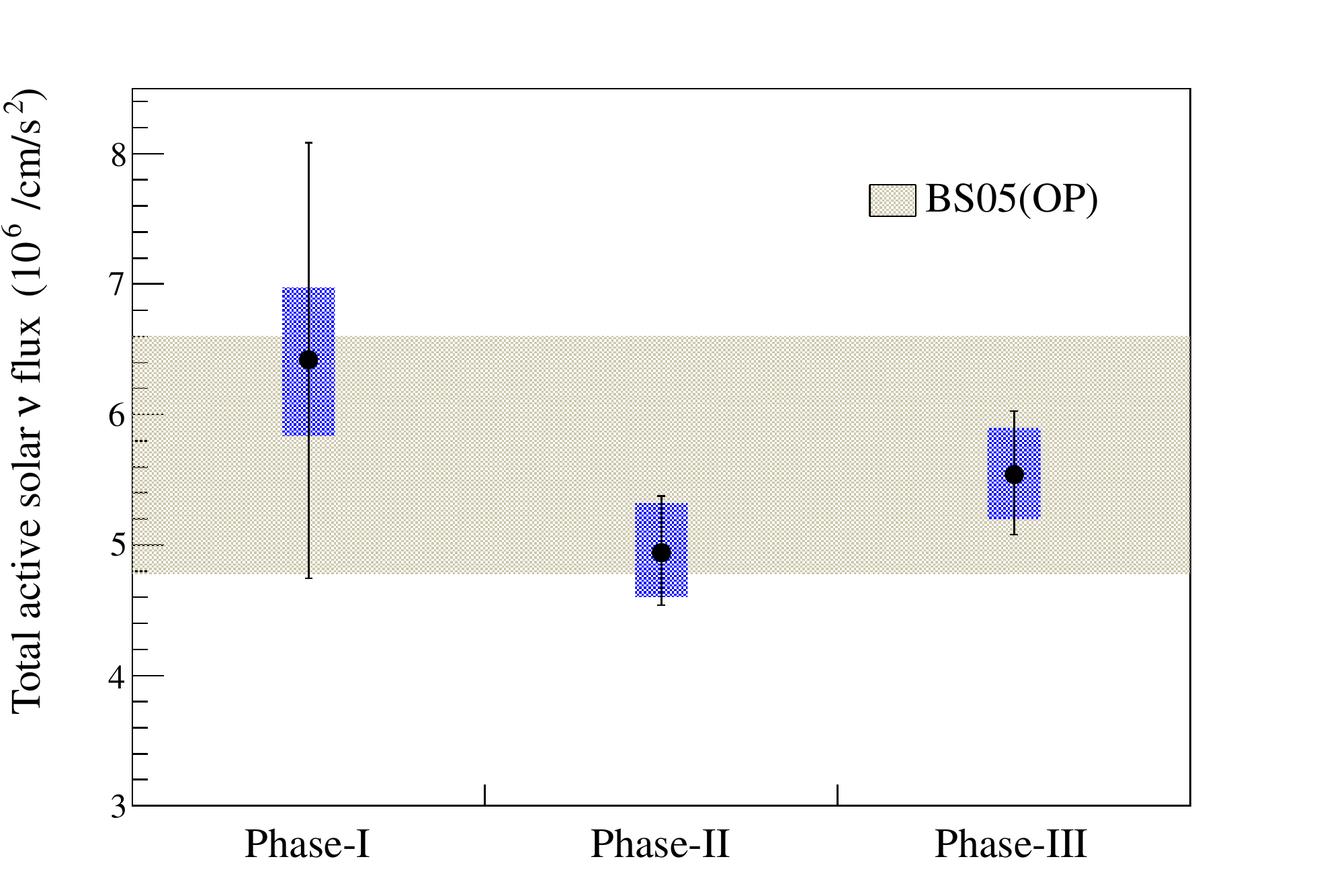}
\caption{\label{fig:nccompare} A comparison of the measured 
energy-unconstrained NC flux results in SNO's three phases.  The 
error bars represent the total $\pm 1\sigma$ uncertainty for the 
measurements, while the height of the shaded boxes represents that 
for the systematic uncertainties.  The horizontal band is the 
1$\sigma$ region of the expected total $^8$B solar neutrino flux in 
the BS05(OP) model~\cite{bs05}.}
\end{center}
\end{figure}

The negative log-likelihood (NLL) function to be minimized was the sum
of a NLL for the PMT array data ($-\log{L_{\textrm{PMT}}}$) and for
the NCD array data ($-\log{L_{\textrm{NCD}}}$).  The spectral
distributions of the ES and CC events were not constrained to the
$^{8}$B shape in the fit, but were extracted from the data.  It should
be noted that the $^8$B spectral shape used here~\cite{wint} differed
from that used in previous phases~\cite{ortiz}.
Figure~\ref{fig:ncdshaper} shows the one-dimensional projection of the
NCD array data overlaid with the best-fit results to signals.  The
energy-unconstrained NC flux results from Phase III are in good
agreement which those in previous phases, as shown in
Fig.~\ref{fig:nccompare}.  It should be emphasized that the 
energy-unconstrained solar neutrino flux measurements are independent 
of solar model inputs.

A detailed description of SNO's Phase-III solar-neutrino measurements
can be found in Refs.~\cite{ncdprl,ncdlongpaper}.

\section{Combined Analysis of all Three Phases}
\label{sec:combined}

The most precise values for the solar neutrino mixing parameters and
the total flux of $^8$B neutrinos from the Sun resulted from a joint
analysis of data from all three phases of the SNO
experiment~\cite{cite:sno3phasePaper}.  The joint analysis accounted
for correlations in systematic uncertainties between phases, and was
based on two distinct strategies. The first was to push toward the
lowest energy threshold possible as it was done in the low-energy
threshold analysis~\cite{bib:leta} described at the end of
Sec.~\ref{sec:sno_ii}, while the second was to strongly leverage the
two independent detection techniques afforded by the combination of
Cherenkov-light data from all three phases and NCD counter data from
Phase III.  The combination of all phases therefore provided a
statistically powerful separation of CC, ES and NC events, and two
independent ways to measure the total flux of active-flavor neutrinos
from $^8$B decay in the Sun.

The data were split into day and night sets in order to search for
matter effects as the neutrinos propagated through the Earth.  The
results of the analysis were presented in the same form as the
low-energy threshold analysis~\cite{bib:leta}, providing the total
\B{} flux, $\fB{}$, independently of any specific active neutrino
flavor oscillation hypothesis; and the energy-dependent \nue{}
survival probability describing the probability that an electron
neutrino remains an electron neutrino in its journey between the Sun
and the SNO detector.  The parameterization of the \B{} signal was
based on an average $\fB{}$ for day and night, a \nue{} survival
probability as a function of neutrino energy, $E_\nu$, during the day,
$\Peed{}$, and an asymmetry between the day and night survival
probabilities, $\Aee{}$. It was defined as
\begin{eqnarray}
\label{eqn:peed}
\Peed{} & = & \Peea{} + \Peeb{} (E_\nu [{\rm MeV}] - 10)\\
\nonumber & &\; + \Peec{} (E_\nu [{\rm MeV}]- 10)^2
\end{eqnarray}
and
\begin{eqnarray}
\Aee{} = 2\frac{\Peen{}-\Peed{}}{\Peen{}+\Peed{}},
\end{eqnarray}
where $\Peen{}$ is the \nue{} survival probability during the night
and with
\begin{eqnarray}
\Aee{} & = & \Aeea{} + \Aeeb{}(E_\nu [{\rm MeV}] - 10).
\label{eqn:aee}
\end{eqnarray}
The parameters $\Aeea{}$, and $\Aeeb{}$ define the relative difference
between the night and day \nue{} survival probability; while
$\Peea{}$, $\Peeb{}$, and $\Peec{}$ define the \nue{} survival
probability during the day. In this parametrization the \nue{}
survival probability during the night is given by
\begin{eqnarray}
\label{eqn:peen}
\Peen{} & = & \Peed \times \frac{1+\Aee{}/2}{1-\Aee{}/2}.
\end{eqnarray}

As with solar neutrino analyses described in previous sections, a
maximum likelihood fit was performed to the Cherenkov events
$\teff{}$, $\rho=(\nicefrac{R}{R_{AV}})^3$, $\be{}$, and $\cts{}$.
The ``shaper energy'', $\Encd{}$, was calculated for each event
recorded with the NCD array.  Monte Carlo simulations assuming the
Standard Solar Model and no neutrino oscillations were used to
determine the event variables for \B{} interactions in the detector.

In the final fit, the events observed in the PMT and NCD arrays were
treated as being uncorrelated, therefore the negative log-likelihood
(NLL) function for all data were given by
\begin{eqnarray}
-\log{L_{\rm data}} = -\log{L_{\rm PMT}} -\log{L_{\rm NCD}},
\end{eqnarray}
where $L_{\rm PMT}$ and $L_{\rm NCD}$, respectively, were the
likelihood functions for the events observed in the PMT and NCD
arrays. The NLL function in the PMT array was given by
\begin{equation}
-\log{L_{\rm PMT}} = \sum_{j=1}^{N}\lambda_{j}(\vec{\eta})
- \sum_{i=1}^{n_{\rm PMT}}\log \left[\sum_{j=1}^{N}\lambda_{j}(\vec{\eta}) f(\vec{x}_i|j,\vec{\eta})\right],
\end{equation}
where $N$ was the number of different event classes, $\vec{\eta}$ was
a vector of ``nuisance'' parameters associated with the systematic
uncertainties, $\lambda_{j}(\vec{\eta})$ was the mean of a Poisson
distribution for the $j^{\rm th}$ class, $\vec{x}_i$ was the vector of
event variables for event $i$, $n_{\rm PMT}$ was the total number of
events in the PMT array during the three phases, and
$f(\vec{x}_i|j,\vec{\eta})$ was the PDF for events of type $j$. The
PDFs for the signal events were re-weighted based on
Eqns.~\ref{eqn:peed} and \ref{eqn:aee}. The NLL function in the NCD
array was given by
\begin{eqnarray}
-\log{L_{\rm NCD}} &= & \frac{1}{2}\left (\frac{\sum_{j=1}^{N}\nu_{j}(\vec{\eta}) - n_{\rm NCD}}{\sigma_{\rm NCD}}\right )^2,
\end{eqnarray}
where $\nu_{j}(\vec{\eta})$ was the mean of a Poisson distribution for
the $j^{\rm th}$ class, $n_{\rm NCD}$ was the total number of neutrons
observed in the NCD array based on the likelihood fit to a histogram
of $\Encd{}$, and $\sigma_{\rm NCD}$ was the associated uncertainty.

\begingroup
\begin{table}[htp]
\centering
\caption{Results from the maximum likelihood fit. Note that $\fB${} is 
in units of $\flux$. The D/N systematic uncertainties include the 
effect of all nuisance parameters that were applied differently 
between day and night. The MC systematic uncertainties include the 
effect of varying the number of events in the Monte Carlo based on 
Poisson statistics. The basic systematic uncertainties include the 
effects of all other nuisance parameters.}
\begin{tabular}{rrrrrrr}	
\hline\hline
&\multicolumn{1}{l}{Best fit}&	Stat.&	\multicolumn{4}{c}{Systematic uncertainty}\\
&			&		&						Basic &					D/N &					MC&						Total\\
\hline
$\fB{}$&		5.25&	$\pm0.16$&				${}^{+0.11}_{-0.12}$&			$\pm0.01$&				${}^{+0.01}_{-0.03}$&		${}^{+0.11}_{-0.13}$\\
$\Peea{}$&	0.317&	$\pm0.016$&				${}^{+0.008}_{-0.010}$&		$\pm0.002$&				${}^{+0.002}_{-0.001}$&		$\pm0.009$\\
$\Peeb{}$&	0.0039&	${}^{+0.0065}_{-0.0067}$&	${}^{+0.0047}_{-0.0038}$&	${}^{+0.0012}_{-0.0018}$&	${}^{+0.0004}_{-0.0008}$&	$\pm0.0045$\\
$\Peec{}$&	-0.0010&	$\pm0.0029$&				${}^{+0.0013}_{-0.0016}$&	${}^{+0.0002}_{-0.0003}$&	${}^{+0.0004}_{-0.0002}$&	${}^{+0.0014}_{-0.0016}$\\
$\Aeea{}$&	0.046&	$\pm0.031$&				${}^{+0.007}_{-0.005}$&		$\pm0.012$&				${}^{+0.002}_{-0.003}$&		${}^{+0.014}_{-0.013}$\\
$\Aeeb{}$&	-0.016&	$\pm0.025$&				${}^{+0.003}_{-0.006}$&		$\pm{0.009}$&				$\pm0.002$&				${}^{+0.010}_{-0.011}$\\
\hline\hline
\end{tabular}
\label{qsigex_ncd_psa_final_3phase_pee}
\end{table}
\endgroup

The final joint fit to all data yielded a total flux of active
neutrino flavors from $\iso{8}{B}$ decays in the Sun of
$\fB{}$=\numberSNOBflux{}. During the day the \nue{} survival
probability at 10\,MeV was given by \numberPeea{}, which was
inconsistent with the null hypothesis that there were no neutrino
oscillations at very high significance.  The results of the combined
fit for $\fB{}$ and the \nue{} survival probability parameters are
summarized in Table~\ref{qsigex_ncd_psa_final_3phase_pee}.  The null
hypothesis that there were no spectral distortions of the \nue{}
survival probability (i.e. $\Peeb{}=0$, $\Peec{}=0$, $\Aeea{}=0$,
$\Aeeb{}=0$), yielded $\Delta\chi^2=1.97$ (26\% C.L) compared to the
best fit. The null hypothesis that there were no day/night distortions
of the \nue{} survival probability (i.e. $\Aeea{}=0$, $\Aeeb{}=0$),
yielded $\Delta\chi^2=1.87$ (61\% C.L.)  compared to the best fit.

\begin{figure}[tbp]
\centering
\includegraphics[width=0.45\columnwidth]{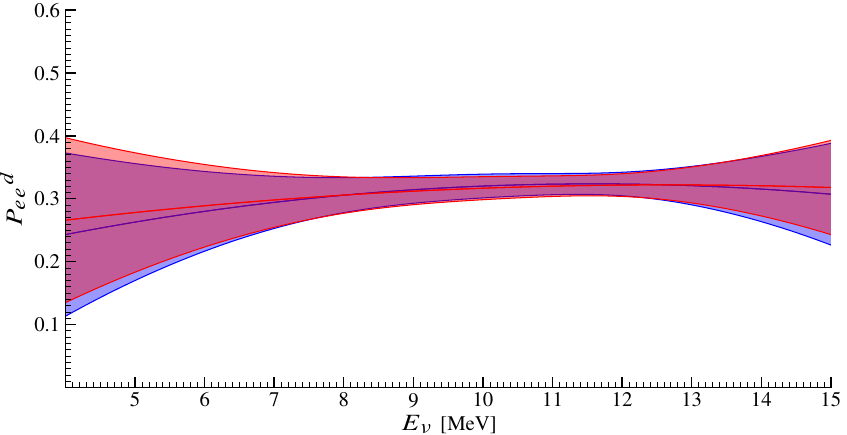}
\includegraphics[width=0.45\columnwidth]{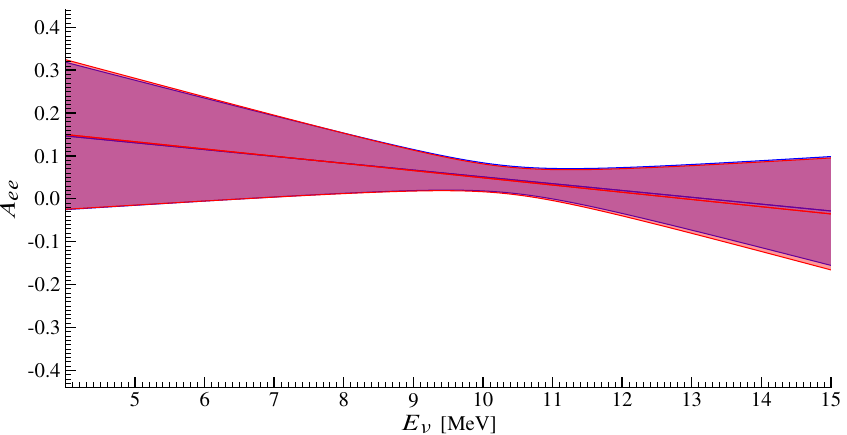}
\caption{Root-mean-square spread in $\Peed{}$ (left) and $\Aee{}$ 
(right), taking into account the parameter uncertainties and 
correlations. The red band represents the results from the maximum 
likelihood fit, and the blue band represents the results from the 
Bayesian fit. The red and blue solid lines, respectively, are the 
best fits from the maximum likelihood and Bayesian fits.}
\label{ncd_psa_final_3phase_pee_cmp}
\end{figure}

Figure~\ref{ncd_psa_final_3phase_pee_cmp} shows the root-mean-square
spread in $\Peed{}$ and $\Aee{}$, taking into account the parameter
uncertainties and correlations. A Bayesian approach was used as
validation analysis and details of this combined analysis are described
in Ref.~\cite{cite:sno3phasePaper}.

\section{Neutrino Oscillations}

The mass differences $\Delta m_{ij}^2$ and the mixing angles $\theta_{ij}$, obtained from neutrino experiments of different source-detector baselines, are used to parametrize the neutrino survival probabilities. Predicting the flux and energy spectrum ($E_\nu$) for all neutrino
flavors requires a model of the neutrino production rates as a
function of location within the Sun, and a model of the survival
probabilities as the neutrinos propagate through the Sun, travel to
the Earth, and then propagate through the Earth. When neutrinos travel through matter, the
survival probabilities are modified due to the Mikheyev-Smirnov-Wolfenstein (MSW) 
effect~\cite{cite:MikheyevSmirnov,cite:Wolfenstein}. For consistency with
previous calculations, the BS05(OP) model~\cite{bs05} was used for the
solar neutrino production rate within the Sun, rather than the more
recent BPS09(GS) or BPS09(AGSS09) models~\cite{ssh:bps09}.  The
$E_\nu$ spectrum for \B{}s was obtained from Ref.~\cite{wint}, while
all other neutrino energy spectra were acquired from
Ref.~\cite{Bahcall}. The electron density as a function of Earth
radius was taken from PREM~\cite{PREM} and
PEM-C~\cite{Dziewonski:1975ih}.

Two different neutrino oscillation hypotheses were considered: 1) the
historical two-flavor neutrino oscillations, which assumed
$\thetaonethree{}=0$ and had two free neutrino oscillation parameters,
$\thetaonetwo{}$ and $\Dmonetwo{}$; and 2) the three-flavor
neutrino oscillations, which fully integrated three free neutrino oscillation
parameters, $\thetaonetwo{}$, $\thetaonethree{}$, and
$\Dmonetwo{}$. The mixing angle, $\thetatwothree{}$, and the
CP-violating phase, $\delta$, are irrelevant for the neutrino
oscillation analysis of solar neutrino data. The solar neutrino data
considered here was insensitive to the exact value $\Dmonethree{}$, so
we used a fixed value of $\pm2.45\times10^{-3}\,{\rm eV^{2}}$ obtained
from long-baseline accelerator experiments and atmospheric neutrino
experiments~\cite{1367-2630-13-6-063004}.  The details of the
oscillation analysis presented here is described in
Ref.~\cite{cite:sno3phasePaper}.


For the two-flavor analysis, Table~\ref{tbl:final:sno} shows the
allowed ranges of the $(\tanthetaonetwo{},\Dmonetwo{})$ parameters
obtained with the SNO results. SNO data alone could not
distinguish between the LMA region and the LOW region, although the
former was slightly favored.  The combination of the SNO results with
the other solar neutrino experimental results eliminated the LOW
region and the higher values of $\Dmonetwo{}$ in the LMA region.
Table~\ref{tbl:final:sno} summarizes the
results from these two-flavor neutrino analyses when the
solar neutrino results were combined with those from the KamLAND (KL)
reactor neutrino experiment~\cite{bib:kl_b8}. 

\begin{figure}	
\centering
\subfigure{\includegraphics[width=0.4\columnwidth]{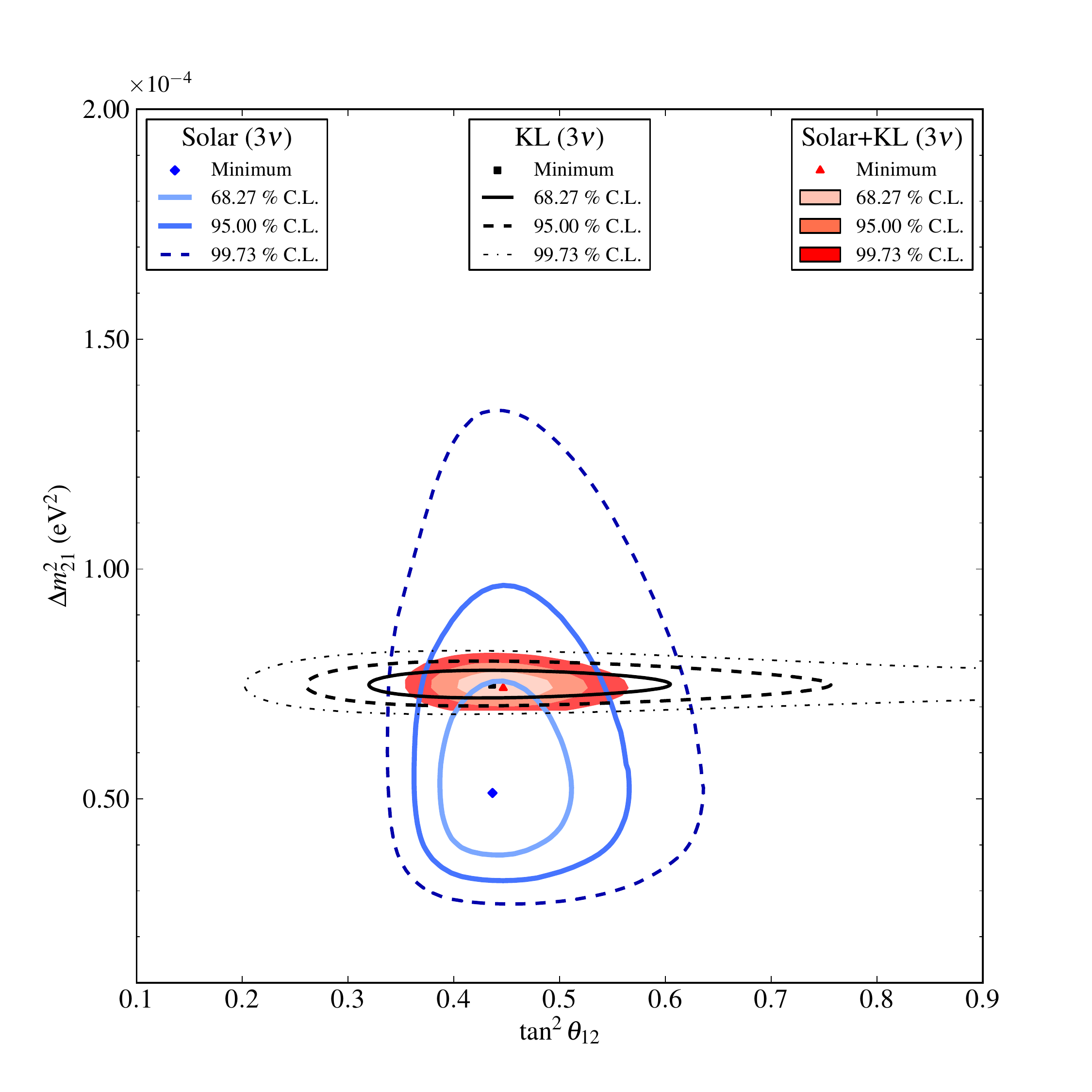}\label{fig:final:global:3nu:12}}
\subfigure{\includegraphics[width=0.4\columnwidth]{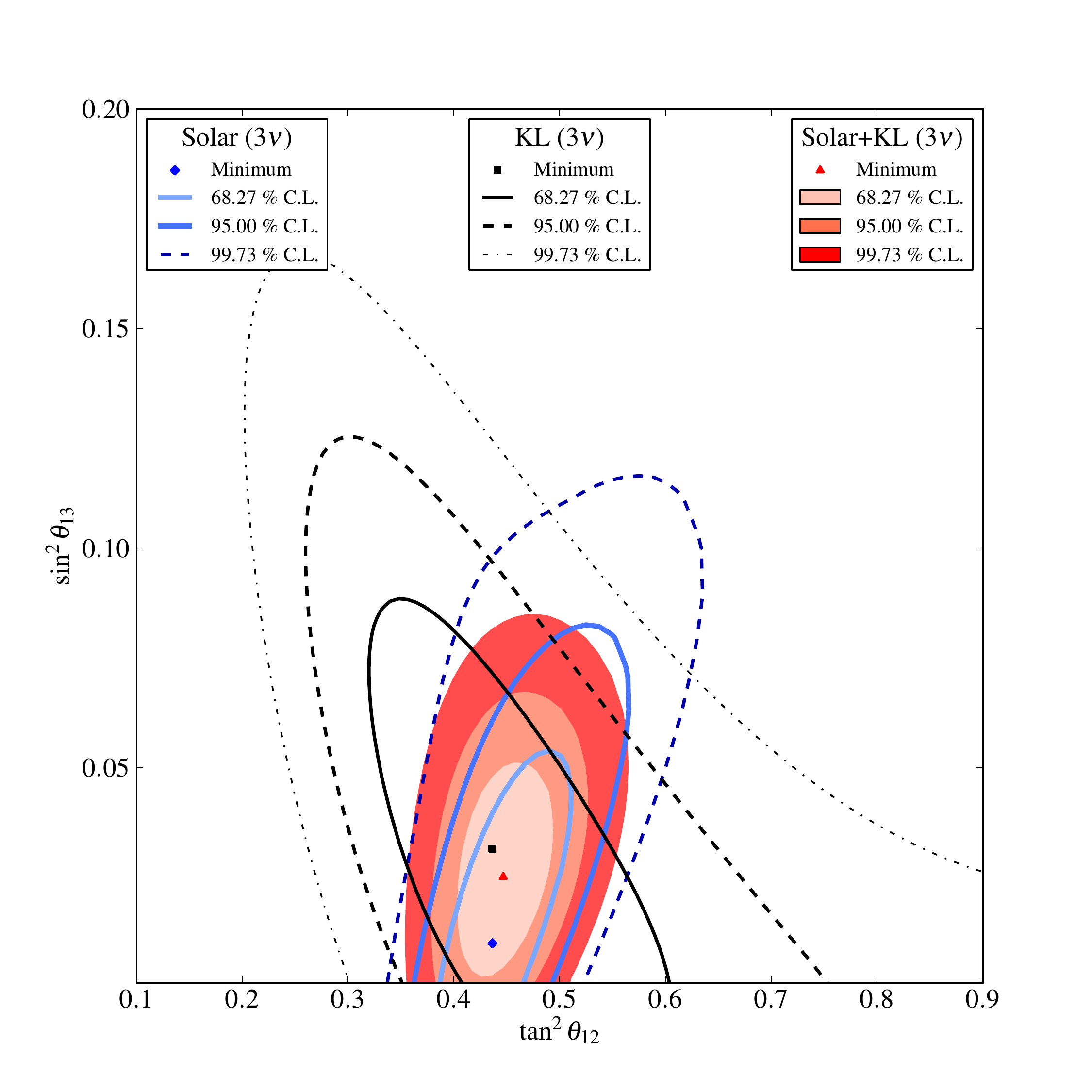}\label{fig:final:global:3nu:13}}
\caption{Three-flavor neutrino oscillation analysis contour using both solar neutrino and KamLAND (KL) results.}
\label{fig:final:global:3nu}
\end{figure}

Figure~\ref{fig:final:global:3nu} shows the allowed regions in the
$(\tanthetaonetwo{}, \Dmonetwo{})$ and $(\tanthetaonetwo{},
\sinthetaonethree{})$ parameter spaces obtained from the results of
all solar neutrino experiments, as well as those including the results
of the KamLAND experiment, in the three-flavor analysis. A non-zero
$\thetaonethree$ has brought the solar neutrino results into better
agreement with the results from the KamLAND experiment.
Table~\ref{tbl:final:global} summarizes the results from these
three-flavor neutrino oscillation analyses. Overall, the
observation by SNO that the average solar \nue{} survival probability at high energy
is about 0.32 and $\theta_{12} \approx 33.5^{\circ}$ corroborate the
matter-induced oscillation scenario of LMA via adiabatic conversion of
electron neutrinos in the core of the Sun.

\begingroup
\begin{table}[htp]
\centering
\caption{Best-fit neutrino oscillation parameters from a two-flavor 
neutrino oscillation analysis. Uncertainties listed are $\pm 
1\sigma$ after the \chis{} was minimized with respect to all other 
parameters.}
\begin{tabular}{lccc}
\hline\hline
Oscillation analysis & $\tanthetaonetwo{}$ & $\Dmonetwo [{\rm eV^{2}}]$ & $\nicefrac{\chi^{2}}{\rm NDF}$\\
\hline
SNO only (LMA)	&$0.427^{+0.033}_{-0.029}$	& $5.62^{+1.92}_{-1.36}\times 10^{-5}$	& $\nicefrac{1.39}{3}$\\
Solar 			&$0.427^{+0.028}_{-0.028}$	& $5.13^{+1.29}_{-0.96}\times 10^{-5}$	& $\nicefrac{108.07}{129}$\\
Solar+KamLAND	&$0.427^{+0.027}_{-0.024}$	& $7.46^{+0.20}_{-0.19}\times 10^{-5}$	& \\
\hline\hline
\end{tabular}
\label{tbl:final:sno}
\end{table}
\endgroup

\begingroup
\begin{table}[htp]
\centering
\caption{Best-fit neutrino oscillation parameters from the three-flavor neutrino 
oscillation analysis in Ref.~\cite{cite:sno3phasePaper}.  Uncertainties 
listed are $\pm 1\sigma$ after the \chis{} was minimized with respect to all 
other parameters. The global analysis includes solar neutrino experiments, 
KamLAND (KL)~\cite{bib:kl_theta13}, and short baseline (SBL) experiments 
(Daya Bay~\cite{cite:dayabay}, RENO~\cite{cite:reno}, 
and Double Chooz~\cite{cite:doublechooz}).}
\begin{tabular}{lcccc}
\hline\hline
Analysis			&$\tanthetaonetwo{}$ & $\Dmonetwo [{\rm eV^{2}}]$ & $\sinthetaonethree (\times 10^{-2})$\\
\hline
Solar			&$0.436^{+0.048}_{-0.036}$	&$5.13^{+1.49}_{-0.98}\times 10^{-5}$	&$<5.8$ (95\% C.L.)\\
Solar+KL			&$0.443^{+0.033}_{-0.026}$	&$7.46^{+0.20}_{-0.19}\times 10^{-5}$	&$2.5^{+1.8}_{-1.4}$\\
				&						&								&$<5.3$ (95\% C.L.)\\
Global (Solar+KL+SBL)			&	$0.443^{+0.030}_{-0.025}$	& $7.46^{+0.20}_{-0.19}\times 10^{-5}$	&$2.49^{+0.20}_{-0.32}$\\
\hline\hline
\end{tabular}
\label{tbl:final:global}
\end{table}
\endgroup

\section{Other Physics Studies}

In addition to the solar neutrino measurements that led to the
discovery of neutrino flavor transformation, the SNO data were also
used to test various aspects of solar models and neutrino properties,
and to search for neutrinos from astrophysical sources.  Neutrinos
from the {\it hep} reaction $^3$H+$p\rightarrow ^4$He$+e^{+}+\nu_e$
has an endpoint energy of 18.77~MeV, but its flux is predicted to be
about three orders of magnitude lower than that of $^8$B neutrinos.
Using the Phase-I data set (0.65~ktons~yr exposure), an upper limit of
$2.3\times 10^4$~cm$^{-2}$s$^{-1}$~(90\% CL) was inferred on the
integral total flux of {\it hep} neutrinos after neutrino oscillations
had been taken into account~\cite{bib:hep_dsnb}.  In the same study, a
search for the diffuse supernova neutrino background (DSNB), which
consists of neutrinos from all extragalactic supernovae since the
formation of stars in the Universe, was performed.  An upper limit of
70~cm$^{-2}$~s$^{-1}$(90\% CL) was found for the $\nu_e$ component of
the DSNB flux in the neutrino energy range of 22.9~MeV$< E_\nu <
$36.9~MeV.  Although this is the most stringent limit on $\nu_e$ flux for 
direct measurements, the Super-Kamiokande experiment has reached an 
upper limit of 2.9~cm$^{-2}$~s$^{-1}$ for the $\bar{\nu}_e$ 
component~\cite{bib:sk_dsnb}.  An analysis to extend these analyses for 
the total three-phase data set is in progress.

The nuclear fusion rate in the solar core should not be affected by
solar rotation or oscillations.  To test this hypothesis, searches on
the periodic variations in $^8$B solar neutrino flux were performed
using Phase-I and Phase-II data sets.  The analysis demonstrated that
the fluctuation of $^8$B neutrino flux was consistent with modulation
by the Earth's orbital eccentricity, and there were no significant
sinusoidal periodicities found with periods between 1~d and
10~years~\cite{bib:b8_periodicity}.  Searches for high-frequency
signals or extra power in the frequency range of 1 to 144~d$^{-1}$ did
not detect any significant signal~\cite{bib:high_freq}.  Additionally
a search in the restricted frequency range of 18.5 to 19.5~d$^{-1}$,
in which ``gravitational-mode'' ($g$-mode) signals had been claimed in
other experiments, did not show any signal.

Although the SNO detector did not observe any large burst of neutrino
events that would be indicative of a galactic supernova explosion, a
thorough study to search for low-multiplicity bursts, defined as
bursts of two or more events that triggered the SNO detector in quick
succession, was performed to look for evidence of distant supernovae
or non-standard supernovae with relatively low neutrino
emission~\cite{bib:low_multiplicity}.  The search had a greater than
50\% detection probability for standard supernovae occurring at a
distance of up to 60~kpc for Phase I and up to 70~kpc for Phase II.
No low-multiplicity bursts were observed.  The correlations of
low-energy signals in the SNO detector and other astrophysical events,
such as gamma-ray bursts and solar flares, were also
studied~\cite{bib:astro_burst}.  No such correlations were found.

The great depth at which the SNO detector was located provided a
unique opportunity to study cosmic-ray and neutrino-induced
through-going muons.  SNO measured the through-going muon flux as a
function of the zenith angles ($\cos\theta_{\rm{zenith}}$), and was
sensitive to neutrino-induced through-going muons in $-1 \leq
\cos\theta_{\rm{zenith}}\leq 0.4$, i.e.~including angles above the
horizon~\cite{bib:muon_flux}.  Total cosmic-ray muon flux at SNO with
$\cos\theta_{\rm{zenith}}>0.4$ was found to be
(3.31$\pm$0.01~(stat.)$\pm$0.09~(syst.))$\times
10^{-10}\mu$/s/cm$^{2}$.  The zenith angle distribution of events
ruled out the case of no neutrino oscillations at the 3$\sigma$ level.
This was the first measurement of the neutrino-induced flux above the
horizon in the angular regime where neutrino oscillations were not an
important effect.

The SNO data were also used to hunt for other new physics.  Using the
data from Phases I and II, SNO was able to constrain the lifetime for
nucleon decay to ``invisible'' modes (such as $n\rightarrow 3\nu$) to
$> 2\times 10^{29}$~y~\cite{bib:invisible_mode} .  This was
accomplished by looking for $\gamma$~rays from the de-excitation of
the residual nucleus that would result from the disappearance of
either a proton or neutron from $^{16}$O.  Non-standard-model physics,
such as spin flavor precession mechanism or neutrino decays, could
potentially convert a small fraction of solar $\nu_e$ to
$\bar{\nu}_e$.  The results from a search for $\bar{\nu}_e$ in
Phase~I~\cite{bib:electron_antinu} confirmed previous results from
similar searches in the Super-Kamiokande~\cite{bib:sk_antinu} and
KamLAND experiments~\cite{bib:kl_antinu}.  An analysis of
$\bar{\nu}_e$ with the full data set is in progress.

\section{Summary}

The principal results from SNO for solar neutrinos show clearly that
electron neutrinos from $^8$B decay in the solar core change their
flavor in transit to Earth. They also provide a measure of the total
flux of $^8$B neutrinos with an accuracy that is better than the
uncertainties in solar models and hopefully will provide guidance in
our detailed understanding of the Sun.  The SNO measurements of the flavor
content of $^8$B solar neutrinos, along with measurements of  
different energy thresholds in other solar neutrino experiment, have 
provided much constraints on $\theta_{12}$, which is unlikely to improve
further until a dedicated medium-baseline reactor neutrino experiment is
online.

\section{Acknowledgements}

This research was supported by: Canada: Natural Sciences and
Engineering Research Council, Industry Canada, National Research
Council, Northern Ontario Heritage Fund, Atomic Energy of Canada,
Ltd., Ontario Power Generation, High Performance Computing Virtual
Laboratory, Canada Foundation for Innovation, Canada Research Chairs;
US: Department of Energy, National Energy Research Scientific
Computing Center, Alfred P. Sloan Foundation; UK: Science and
Technology Facilities Council; Portugal: Funda\c{c}\~{a}o para a
Ci\^{e}ncia e a Tecnologia. We thank the SNO technical staff for their
strong contributions. We thank Vale (formerly Inco, Ltd.) for hosting
this project.

\bibliographystyle{elsarticle-num}

\begin{thebibliography}{999}
\expandafter\ifx\csname natexlab\endcsname\relax\def\natexlab#1{#1}\fi
\expandafter\ifx\csname bibnamefont\endcsname\relax
\def\bibnamefont#1{#1}\fi
\expandafter\ifx\csname bibfnamefont\endcsname\relax
\def\bibfnamefont#1{#1}\fi
\expandafter\ifx\csname citenamefont\endcsname\relax
\def\citenamefont#1{#1}\fi
\expandafter\ifx\csname url\endcsname\relax
\def\url#1{\texttt{#1}}\fi
\expandafter\ifx\csname urlprefix\endcsname\relax\def\urlprefix{URL
}\fi
\providecommand{\bibinfo}[2]{#2}
\providecommand{\eprint}[2][]{\url{#2}}

\bibitem{jnb_na}
\bibinfo{author}{\bibnamefont{J.N. Bahcall}},
\bibinfo{journal}{Neutrino Astrophysics},
\bibinfo{publisher}{Cambridge University Press} (\bibinfo{year}{1989}).

\bibitem{chen}
\bibinfo{author}{\bibnamefont{{H.~H.~Chen}}},
\bibinfo{journal}{Phys. Rev. Lett.} \textbf{\bibinfo{volume}{55}},
\bibinfo{pages}{1534} (\bibinfo{year}{1985}).

\bibitem{sinclairINC}
\bibinfo{author}{\bibnamefont{{D.~Sinclair, A.L.~Carter, D.~Kessler, E.D.~Earle, P.~Jagam, J.J.~Simpson, R.C.~Allen, H.H.~Chen, P.J.~Doe, E.D.~Hallman, W.F.~Davidson, A.B.~McDonald, R.S.~Storey, G.T.~Ewan, H.B.~Mak, B.C.~Robertson}}},
\bibinfo{journal}{Il Nuovo Cimento C} \textbf{\bibinfo{volume}{9}},
\bibinfo{pages}{308} (\bibinfo{year}{1986}).

\bibitem{bp2000} 
\bibinfo{author}{\bibnamefont{J.N. Bahcall, M.H. Pinsonneault, and S. Basu}}, 
\bibinfo{journal}{Astrophys. J.} \textbf{\bibinfo{volume}{555}}, 
\bibinfo{pages}{990}(\bibinfo{year}{2001}).

\bibitem{NIM} 
\bibinfo{author}{\bibnamefont{{J. Boger {\emph{et~al.}} (SNO Collaboration)}}}, 
\bibinfo{journal}{Nucl. Instr. and Meth. A} \textbf{\bibinfo{volume}{449}}, 
\bibinfo{pages}{172}(\bibinfo{year}{2000}).

\bibitem{bib:n16} 
\bibinfo{author}{\bibnamefont{{M.R. Dragowsky~\textit{et al.}}}}, 
\bibinfo{journal}{Nucl. Inst. Meth. A} \textbf{\bibinfo{volume}{481}}, 
\bibinfo{pages}{284} (\bibinfo{year}{2002}).

\bibitem{bib:li8} 
\bibinfo{author}{\bibnamefont{{N.J. Tagg {\emph{et~al.}}}}}, 
\bibinfo{journal}{Nucl. Instr. and Meth. A} \textbf{\bibinfo{volume}{489}}, 
\bibinfo{pages}{178}(\bibinfo{year}{2002}).

\bibitem{bib:pt} 
\bibinfo{author}{\bibnamefont{{A.W.P. Poon {\emph{et~al.}}}}}, 
\bibinfo{journal}{Nucl. Instr. and Meth. A} \textbf{\bibinfo{volume}{452}}, 
\bibinfo{pages}{115}(\bibinfo{year}{2000}).

\bibitem[{\citenamefont{{T.~C.~Andersen~{\emph{et~al.}}}}(2003{\natexlab{a}})}]{bib:mnox}
\bibinfo{author}{\bibnamefont{{T.~C.~Andersen~{\emph{et~al.}}}}},
  \bibinfo{journal}{Nucl. Instr. and Meth. A} \textbf{\bibinfo{volume}{501}},
  \bibinfo{pages}{399} (\bibinfo{year}{2003}{\natexlab{a}}).

\bibitem[{\citenamefont{{T.~C.~Andersen~{\emph{et~al.}}}}(2003{\natexlab{b}})}]{bib:htio}
\bibinfo{author}{\bibnamefont{{T.~C.~Andersen~{\emph{et~al.}}}}},
  \bibinfo{journal}{Nucl. Instr. and Meth. A} \textbf{\bibinfo{volume}{501}},
  \bibinfo{pages}{386} (\bibinfo{year}{2003}{\natexlab{b}}).

\bibitem{longd2o}
\bibinfo{author}{\bibnamefont{{B. Aharmim {\emph {et~al.}} (SNO Collaboration)}}}, \bibinfo{journal}{Phys. Rev. C} \textbf{\bibinfo{volume}{75}}, \bibinfo{pages}{045502}
  (\bibinfo{year}{2007}{\natexlab{a}}).

\bibitem{laserball} 
\bibinfo{author}{\bibnamefont{{B. A. Moffat~\textit{et al.}}}}, 
\bibinfo{journal}{Nucl. Inst. Meth. A} \textbf{\bibinfo{volume}{554}}, 
\bibinfo{pages}{255} (\bibinfo{year}{2005}).

\bibitem{snocc}
\bibinfo{author}{\bibnamefont{Q. R. Ahmad {\emph {et~al.}} (SNO Collaboration)}}, \bibinfo{journal}{Phys.
  Rev. Lett.} \textbf{\bibinfo{volume}{87}}, \bibinfo{pages}{071301}
  (\bibinfo{year}{2001}).
  
\bibitem{snonc}
\bibinfo{author}{\bibnamefont{Q. R. Ahmad {\emph {et~al.}} (SNO Collaboration)}}, \bibinfo{journal}{Phys.
  Rev. Lett.} \textbf{\bibinfo{volume}{89}}, \bibinfo{pages}{011301}
  (\bibinfo{year}{2002}{\natexlab{a}}).

\bibitem{TC} 
\bibinfo{author}{\bibnamefont{A.S. Brun, S. Turck-Chi\`{e}ze, and J.P. Zahn}}, 
\bibinfo{journal}{Astrophys. J.} \textbf{\bibinfo{volume}{525}}, 
\bibinfo{pages}{1032}(\bibinfo{year}{2001}).

\bibitem{bib:d2odaynight}
\bibinfo{author}{\bibnamefont{{Q.R. Ahmad {\emph {et~al.}} (SNO Collaboration)}}}, 
\bibinfo{journal}{Phys. Rev. Lett.} \textbf{\bibinfo{volume}{89}}, 
\bibinfo{pages}{011302} (\bibinfo{year}{2002}{\natexlab{a}}).

\bibitem{saltprl}
\bibinfo{author}{\bibnamefont{{S.N. Ahmed {\emph {et~al.}} (SNO Collaboration)}}}, 
\bibinfo{journal}{Phys. Rev. Lett.} \textbf{\bibinfo{volume}{92}}, 
\bibinfo{pages}{181301} (\bibinfo{year}{2004}{\natexlab{a}}).

\bibitem{nsp}
\bibinfo{author}{\bibnamefont{{B. Aharmim {\emph {et~al.}} (SNO Collaboration)}}}, 
\bibinfo{journal}{Phys. Rev. C} \textbf{\bibinfo{volume}{72}}, 
\bibinfo{pages}{055502} (\bibinfo{year}{2005}{\natexlab{a}}).

\bibitem{bib:leta}
\bibinfo{author}{\bibnamefont{{B. Aharmim {\textit{et~al.}} (SNO Collaboration)}}}, 
\bibinfo{journal}{Phys. Rev. C} \textbf{\bibinfo{volume}{81}}, \bibinfo{pages}{055504} (\bibinfo{year}{2010}{\natexlab{a}}).

\bibitem{ncdcountersnim} 
\bibinfo{author}{\bibnamefont{{J. Amsbaugh \textit{et al.}}}}, 
\bibinfo{journal}{Nucl. Inst. Meth. A} \textbf{\bibinfo{volume}{579}}, 
\bibinfo{pages}{1054} (\bibinfo{year}{2007}).

\bibitem{na24_nim_paper}
\bibinfo{author}{\bibnamefont{{K.~Boudjemline {\textit{et~al.}}}}},
\bibinfo{journal}{Nucl. Instr. Meth. A}
\textbf{\bibinfo{volume}{620}}, 
\bibinfo{pages}{171} (\bibinfo{year}{2010}).

\bibitem{HSNIM} 
\bibinfo{author}{\bibnamefont{{H. M. O'Keeffe {\textit{et~al.}}}}},
\bibinfo{journal}{Nucl. Inst. Meth. A} \textbf{\bibinfo{volume}{659}}, 
\bibinfo{pages}{182} (\bibinfo{year}{2011}).

\bibitem{ncdsim} 
\bibinfo{author}{\bibnamefont{{B.~Beltran~{\textit{et~al.}}}}},
\bibinfo{journal}{New J. Phys.} \textbf{\bibinfo{volume}{13}},
\bibinfo{pages}{073006} (\bibinfo{year}{2011}{\natexlab{a}}).

\bibitem{wint} 
\bibinfo{author}{\bibnamefont{W.T.~Winter~\textit{et al.}}}, 
\bibinfo{journal}{Phys. Rev. C} \textbf{\bibinfo{volume}{73}}, 
\bibinfo{pages}{025503} (\bibinfo{year}{2006}).

\bibitem[{\citenamefont{{C.E. Ortiz~{\textit{et~al.}}}}()}]{ortiz}
\bibinfo{author}{\bibnamefont{{C.E. Ortiz~{\textit{et~al.}}}}},
\bibinfo{journal}{Phys. Rev. Lett.} \textbf{\bibinfo{volume}{85}}, \bibinfo{pages}{2909}
  (\bibinfo{year}{2000}).

\bibitem[{\citenamefont{Bahcall et~al.}(2001)\citenamefont{Bahcall,
  Serenelli, and Basu}}]{bs05}
\bibinfo{author}{\bibfnamefont{J.~N.} \bibnamefont{Bahcall}},
  \bibinfo{author}{\bibfnamefont{A.~M.}~\bibnamefont{Serenelli}},
  \bibnamefont{and} \bibinfo{author}{\bibfnamefont{S.}~\bibnamefont{Basu}},
  \bibinfo{journal}{Astrophys. J.} \textbf{\bibinfo{volume}{621}},
  \bibinfo{pages}{L85} (\bibinfo{year}{2005}).

\bibitem{ncdprl}
\bibinfo{author}{\bibnamefont{{B. Aharmim {\textit{et~al.}} (SNO Collaboration)}}}, 
\bibinfo{journal}{Phys. Rev. Lett.} \textbf{\bibinfo{volume}{101}}, \bibinfo{pages}{111301} (\bibinfo{year}{2008}{\natexlab{a}}).

\bibitem{ncdlongpaper}
\bibinfo{author}{\bibnamefont{{B. Aharmim {\textit{et~al.}} (SNO Collaboration)}}}, 
\bibinfo{journal}{Phys. Rev. C} \textbf{\bibinfo{volume}{87}}, \bibinfo{pages}{015502} (\bibinfo{year}{2013}{\natexlab{a}}).

\bibitem [{\citenamefont {Aharmim}\ \emph {et~al.}(2010)\citenamefont {Aharmim}
  \emph {et~al.}}]{cite:sno3phasePaper}%
  \bibfield  {author} {\bibinfo {author} {\bibfnamefont {B.}~\bibnamefont
  {Aharmim}} \emph {et~al.} (\bibinfo {collaboration} {SNO Collaboration}),\
  }{\bibfield  {journal} {\bibinfo
  {journal} {Phys. Rev. C}\ }\textbf {\bibinfo {volume} {88}},\ \bibinfo
  {pages} {025501} (\bibinfo {year} {2013}).}
%
\bibitem{cite:MikheyevSmirnov}
\bibinfo{author}{\bibfnamefont{S.~P.}\ \bibnamefont{Mikheyev}}\ and\  {\bibfnamefont{A.~Y.}\ \bibnamefont{Smirnov}},
\bibinfo{journal}{Nuovo Cim. C} \textbf{\bibinfo{volume}{9}}, 
\bibinfo{pages}{17} (\bibinfo{year}{1986}{\natexlab{a}}).
%
\bibitem{cite:Wolfenstein}
\bibinfo{author}{\bibfnamefont{L.}\ \bibnamefont{Wolfenstein}},
\bibinfo{journal}{Phys. Rev. D} \textbf{\bibinfo{volume}{17}}, 
\bibinfo{pages}{2369} (\bibinfo{year}{1978}{\natexlab{a}}).
%
\bibitem [{\citenamefont {Serenelli}\ \emph {et~al.}(2009)\citenamefont
  {Serenelli}, \citenamefont {Basu}, \citenamefont {Ferguson},\ and\
  \citenamefont {Asplund}}]{ssh:bps09}%
  \bibfield  {author} {\bibinfo {author} {\bibfnamefont {A.~M.}\ \bibnamefont
  {Serenelli}}, \bibinfo {author} {\bibfnamefont {S.}~\bibnamefont {Basu}},
  \bibinfo {author} {\bibfnamefont {J.~W.}\ \bibnamefont {Ferguson}}, \ and\
  \bibinfo {author} {\bibfnamefont {M.}~\bibnamefont {Asplund}},\ }
  {\bibfield  {journal} {\bibinfo
  {journal} {Astrophys. J. Lett.}\ }\textbf {\bibinfo {volume} {705}},\
  \bibinfo {pages} {L123} (\bibinfo {year} {2009}).}%
%
\bibitem [{Bah()}]{Bahcall}%
  \url {http://www.sns.ias.edu/$\sim$jnb/SNdata/sndata.html}
%
\bibitem [{\citenamefont {{Dziewonski}}\ and\ \citenamefont
  {{Anderson}}(1981)}]{PREM}%
  \bibfield  {author} {\bibinfo {author} {\bibfnamefont {A.~M.}\ \bibnamefont
  {{Dziewonski}}}\ and\ \bibinfo {author} {\bibfnamefont {D.~L.}\ \bibnamefont
  {{Anderson}}},\ } {\bibfield
  {journal} {\bibinfo  {journal} {Phys. Earth Planet. In.}\ }\textbf {\bibinfo
  {volume} {25}},\ \bibinfo {pages} {297} (\bibinfo {year} {1981}).}
%
\bibitem [{\citenamefont {Dziewonski}\ \emph {et~al.}(1975).\citenamefont
  {Dziewonski}, \citenamefont {Hales},\ and\ \citenamefont
  {Lapwood}}]{Dziewonski:1975ih}%
  \bibfield  {author} {\bibinfo {author} {\bibfnamefont {A.~M.}\ \bibnamefont
  {Dziewonski}}, \bibinfo {author} {\bibfnamefont {A.~L.}\ \bibnamefont
  {Hales}}, \ and\ \bibinfo {author} {\bibfnamefont {E.~R.}\ \bibnamefont
  {Lapwood}},\ } {\bibfield
  {journal} {\bibinfo  {journal} {Phys. Earth Planet. In.}\ }\textbf {\bibinfo
  {volume} {10}},\ \bibinfo {pages} {12 } (\bibinfo {year} {1975}).}
%
\bibitem [{\citenamefont {Schwetz}\ \emph {et~al.}(2011)\citenamefont
  {Schwetz}, \citenamefont {T\'{o}rtola},\ and\ \citenamefont
  {Valle}}]{1367-2630-13-6-063004}%
  \bibfield  {author} {\bibinfo {author} {\bibfnamefont {T.}~\bibnamefont
  {Schwetz}}, \bibinfo {author} {\bibfnamefont {M.}~\bibnamefont
  {T\'{o}rtola}}, \ and\ \bibinfo {author} {\bibfnamefont {J.~W.~F.}\
  \bibnamefont {Valle}},\ }
  {\bibfield  {journal} {\bibinfo  {journal} {New J. Phys.}\ }\textbf {\bibinfo
  {volume} {13}},\ \bibinfo {pages} {063004} (\bibinfo {year}
  {2011})}
%
\bibitem{bib:kl_b8}
\bibinfo{author}{\bibnamefont{{S. Abe {\emph {et~al.}} (KamLAND Collaboration)}}},
\bibinfo{journal}{Phys. Rev. C} \textbf{\bibinfo{volume}{84}}, 
\bibinfo{pages}{035804} (\bibinfo{year}{2011}{\natexlab{a}}).
%
\bibitem{bib:kl_theta13}
\bibinfo{author}{\bibnamefont{{A. Gando {\emph {et~al.}} (KamLAND Collaboration)}}} 
\bibinfo{journal}{Phys. Rev. D} \textbf{\bibinfo{volume}{83}}, 
\bibinfo{pages}{052002} (\bibinfo{year}{2011}{\natexlab{a}}).
%
\bibitem{cite:dayabay}
\bibinfo{author}{\bibnamefont{{F.P.~An {\textit{et~al.}} (Daya Bay Collaboration)}}}, 
\bibinfo{journal}{Chinese Physics C} \textbf{\bibinfo{volume}{37}}, 
\bibinfo{pages}{011001} (\bibinfo{year}{2013}{\natexlab{a}}).
%
\bibitem{cite:reno}
\bibinfo{author}{\bibnamefont{{J.K.~Ahn {\textit{et~al.}} (RENO Collaboration)}}}, 
\bibinfo{journal}{Phys. Rev. Lett.} \textbf{\bibinfo{volume}{108}}, 
\bibinfo{pages}{191802} (\bibinfo{year}{2012}{\natexlab{a}}).
%
\bibitem{cite:doublechooz}
\bibinfo{author}{\bibnamefont{{Y.~Abe {\textit{et~al.}} (Double Chooz Collaboration)}}}, 
\bibinfo{journal}{Phys. Rev. Lett.} \textbf{\bibinfo{volume}{108}}, 
\bibinfo{pages}{131801} (\bibinfo{year}{2013}{\natexlab{a}}).

\bibitem{bib:hep_dsnb}
\bibinfo{author}{\bibnamefont{{B. Aharmim {\emph {et~al.}} (SNO Collaboration)}}}, 
\bibinfo{journal}{Astrophys. J.} \textbf{\bibinfo{volume}{653}}, 
\bibinfo{pages}{1545} (\bibinfo{year}{2006}{\natexlab{a}}).

\bibitem{bib:sk_dsnb}
\bibinfo{author}{\bibnamefont{{K. Bays {\emph {et~al.}} (Super-Kamiokande Collaboration)}}}, 
\bibinfo{journal}{Phys. Rev. D} \textbf{\bibinfo{volume}{85}}, 
\bibinfo{pages}{052007} (\bibinfo{year}{2012}{\natexlab{a}}).

\bibitem{bib:b8_periodicity}
\bibinfo{author}{\bibnamefont{{B. Aharmim {\emph {et~al.}} (SNO Collaboration)}}}, 
\bibinfo{journal}{Phys. Rev. D} \textbf{\bibinfo{volume}{72}}, 
\bibinfo{pages}{052010} (\bibinfo{year}{2005}{\natexlab{a}}).

\bibitem{bib:high_freq}
\bibinfo{author}{\bibnamefont{{B. Aharmim {\emph {et~al.}} (SNO Collaboration)}}}, 
\bibinfo{journal}{Astrophys. J.} \textbf{\bibinfo{volume}{710}}, 
\bibinfo{pages}{540} (\bibinfo{year}{2010}{\natexlab{a}}).

\bibitem{bib:low_multiplicity}
\bibinfo{author}{\bibnamefont{{B. Aharmim {\emph {et~al.}} (SNO Collaboration)}}}, 
\bibinfo{journal}{Astrophys. J.} \textbf{\bibinfo{volume}{728}}, 
\bibinfo{pages}{83} (\bibinfo{year}{2011}{\natexlab{a}}).

\bibitem{bib:astro_burst}
\bibinfo{author}{\bibnamefont{{B. Aharmim {\emph {et~al.}} (SNO Collaboration)}}}, 
\bibinfo{journal}{Astropart. Phys.} \textbf{\bibinfo{volume}{55}}, 
\bibinfo{pages}{1} (\bibinfo{year}{2014}{\natexlab{a}}).

\bibitem{bib:muon_flux}
\bibinfo{author}{\bibnamefont{{B. Aharmim {\emph {et~al.}} (SNO Collaboration)}}}, 
\bibinfo{journal}{Phys. Rev. D} \textbf{\bibinfo{volume}{80}}, 
\bibinfo{pages}{012001} (\bibinfo{year}{2009}{\natexlab{a}}).

\bibitem{bib:invisible_mode}
\bibinfo{author}{\bibnamefont{{S.N. Ahmed {\emph {et~al.}} (SNO Collaboration)}}}, 
\bibinfo{journal}{Phys. Rev. Lett.} \textbf{\bibinfo{volume}{92}}, 
\bibinfo{pages}{102004} (\bibinfo{year}{2004}{\natexlab{a}}).

\bibitem{bib:electron_antinu}
\bibinfo{author}{\bibnamefont{{B. Aharmim {\emph {et~al.}} (SNO Collaboration)}}}, 
\bibinfo{journal}{Phys. Rev. D} \textbf{\bibinfo{volume}{70}}, 
\bibinfo{pages}{093014} (\bibinfo{year}{2004}{\natexlab{a}}).

\bibitem{bib:sk_antinu}
\bibinfo{author}{\bibnamefont{{Y. Gando {\emph {et~al.}} (Super-Kamiokande Collaboration)}}}, 
\bibinfo{journal}{Phys. Rev. Lett.} \textbf{\bibinfo{volume}{90}}, 
\bibinfo{pages}{171302} (\bibinfo{year}{2003}{\natexlab{a}}).

\bibitem{bib:kl_antinu}
\bibinfo{author}{\bibnamefont{{K. Eguchi {\emph {et~al.}} (KamLAND Collaboration)}}},
\bibinfo{journal}{Phys. Rev. Lett.} \textbf{\bibinfo{volume}{92}}, 
\bibinfo{pages}{071301} (\bibinfo{year}{2004}{\natexlab{a}}).
%

\end{thebibliography}

\end{document}